\documentclass[aps,preprint,nofootinbib,floatfix]{revtex4-1}
\pdfoutput=1
\usepackage{amsmath,amssymb}
\usepackage{graphicx}
\usepackage{url,hyperref}

\usepackage{epsfig}
\def\be{\begin{equation}}
\def\ee{\end{equation}}
\def\bea{\begin{eqnarray}}
\def\eea{\end{eqnarray}}

\catcode`\@=11
\def\lsim{\mathrel{\mathpalette\@versim<}}
\def\gsim{\mathrel{\mathpalette\@versim>}}
\def\@versim#1#2{\vcenter{\offinterlineskip
\ialign{$\m@th#1\hfil##\hfil$\crcr#2\crcr\sim\crcr } }}
\catcode`\@=12
\usepackage{axodraw}

\catcode`@=12
\begin{document}
\thispagestyle{empty}
\begin{flushright}
UCRHEP-T536\\
\end{flushright}
\begin{center}
{\LARGE \bf Multilepton Higgs Decays\\ through the 
Dark Portal\\}
\vspace{1.2in}
{\bf Chia-Feng Chang$^1$, Ernest Ma$^2$ and Tzu-Chiang Yuan$^3$\\}
\vspace{0.2in}
{\sl $^1$ Department of Physics, National Taiwan Normal University, Taipei 116, Taiwan\\}
\vspace{0.1in}
{\sl $^2$ Department of Physics and Astronomy, University of California,\\
Riverside, California 92521, USA\\}
\vspace{0.1in}
{\sl $^3$ Institute of Physics, Academia Sinica, 
Nangang, Taipei 11529, Taiwan\\}
\end{center}
\vspace{1in}
\begin{abstract}
The $U(1)_D$ gauge sector containing one dark Higgs boson $h_D$ and one 
dark photon $\gamma_D$ may be explored through the decays of the 126 GeV 
particle discovered at the Large Hadron Collider (LHC), assumed here as the 
heavier mass eigenstate $h_1$ in the mixing of the standard model $h$ with 
$h_D$.  The various decays of $h_1$ to $\gamma_D \gamma_D$, $h_2 h_2$, 
$h_2 \gamma_D \gamma_D$ and $h_2 h_2 h_2$ would yield multilepton final 
states through the mixing of $\gamma_D$ with the photon and the decay 
$h_2 \to \gamma_D \gamma_D$, where $h_2$ is the lighter dark Higgs. 
Future searches for signals of multilepton jets at the LHC may reveal 
the existence of this possible dark sector governed simply by the original Abelian Higgs model.
\end{abstract}

\maketitle



\section{Introduction}

The original Higgs model \cite{higgs} of spontaneous symmetry breaking involves just one 
complex scalar field $\chi$ and one vector gauge field $C$.  As $\chi$ 
acquires a nonzero vacuum expectation value (VEV), the physical spectrum of this 
model consists of a massive vector boson $\gamma_D$ and a massive real scalar 
boson $h_D$, and the only interactions between them are of the form 
$h_D \gamma_D \gamma_D$ and $h_D^2 \gamma_D \gamma_D$.  The analog of $h_D$ 
in the electroweak 
$SU(2) \times U(1)$ extension \cite{SM} of this original model, commonly called the 
Higgs boson $h$, is presumably the 126 GeV particle observed at the Large 
Hadron Collider (LHC) \cite{ATLAS,CMS}.  Is this the whole story?  Perhaps not, because 
the original Higgs model may still be realized physically, but in a sector 
which connects with the standard model (SM) of particle interactions only 
through $h_D - h$ mass mixing and $\gamma_D - \gamma$ kinetic mixing \cite{Holdom}. 
If so, the 126 GeV particle may be identified with the heavier mass 
eigenstate $h_1$ and decays such as $h_1$ to $\gamma_D \gamma_D$, $h_2 h_2$, $h_2 \gamma_D \gamma_D$
and $h_2 h_2 h_2$ would result in multilepton final states
via $\gamma_D \to \bar l l$ or $h_2 \to \gamma_D \gamma_D$ and 
then followed by $\gamma_D \to \bar l l$, where $h_2$ is the lighter dark Higgs and $l$ is the SM lepton.

In Sec.~2 we set up our model.  
Phenomenology based on similar model has been studied before, see for example 
Refs.~\cite{Gopalakrishna:2008dv,Davoudiasl-etal-1,Davoudiasl-etal-2,Zupan-1,Zupan-2,Feldman-Liu-Nath} 
and references therein.
In Sec.~3 we consider mixing effects in the scalar sector as well as the gauge boson sector.
We show the $h_D-h$ mixing in 
detail and present all the relevant trilinear and quadrilinear couplings 
of the physical $h_1$ and $h_2$ bosons.  
We also briefly discuss the mixings between the three neutral gauge bosons in the model
as studied previously in Ref.~\cite{Feldman-Liu-Nath}.
In Sec.~4 we discuss the possible decay modes of the SM Higgs outside those of the SM and their several 
kinematic regions. In Sec.~5 we present numerical results for various branching ratios of the non-standard 
decay modes of the SM Higgs, identified here as $h_1$.
In Sec.~6 we study the signals of multilepton jets of the model at the LHC-14. We conclude in Sec.~7.


\section{$SU(2)_L \times U(1)_Y \times U(1)_D$ Model}

We extend the electroweak SM by including the original Abelian Higgs model for a dark $U(1)_D$~\cite{Gopalakrishna:2008dv,Davoudiasl-etal-1,Davoudiasl-etal-2,Zupan-1,Zupan-2,Feldman-Liu-Nath}.
The bosonic part of the Lagrangian density is
\begin{equation}
{\cal L}_{\rm B} = {\cal L}_{\rm gauge} + {\cal L}_{\rm scalar} 
\end{equation}
with
\begin{equation}
\label{Lgauge}
{\cal L}_{\rm gauge} = -\frac{1}{4} {\vec W}_{\mu \nu} \cdot {\vec W}^{\mu \nu}
-\frac{1}{4} B_{\mu \nu} B^{\mu \nu}
-\frac{1}{4} C_{\mu \nu} C^{\mu \nu}
-\frac{\epsilon}{2} B_{\mu \nu} C^{\mu \nu} \; ,
\end{equation}
\begin{equation}
\label{Lscalar}
{\cal L}_{\rm scalar} = \vert D_\mu \Phi \vert^2 + \vert D_\mu \chi \vert^2 - V _{\rm scalar}(\Phi, \chi) \; ,
\end{equation}
and
\begin{eqnarray}
\label{covd}
D_\mu \Phi & = & \left( \partial_\mu + i g \frac{1}{2}\sigma_a W_{a \mu} + i  \frac{1}{2} g' B_\mu \right) \Phi \; ,\\ 
D_\mu \chi & = & 
\left( \partial_\mu + i g_D C_\mu \right) \chi \; ,
\end{eqnarray}
where ${\vec W}^\mu$, $B^\mu$ and $C^\mu$ are the gauge potentials of the 
$SU(2)_L$, $U(1)_Y$ and $U(1)_D$ with gauge couplings $g$, $g'$ and $g_D$ respectively,
and $\epsilon$ is the kinetic mixing parameter between the two $U(1)$s \cite{Holdom}.
The scalar potential in (\ref{Lscalar}) is given by 
\begin{equation}
\label{scalarpot}
V_{\rm scalar}  =  - \mu_\Phi^2 \Phi^\dagger \Phi +  \lambda_\Phi \left( \Phi^\dagger \Phi \right)^2 
- \mu^2_{\chi} \chi^* \chi  + \lambda_{\chi} \left( \chi^* \chi \right)^2 
+  \lambda_{\Phi\chi} \left( \Phi^\dagger \Phi \right) \left( \chi^* \chi  \right) \; .
\end{equation}
We pick the unitary gauge and expand the scalar fields around the vacuum
\begin{equation}
\Phi (x) = \frac{1}{\sqrt 2} 
\left(
\begin{tabular}{c}
0
\\
$v + h(x)$
\end{tabular}
\right)
\;\;\; , \;\;\; 
\chi (x) = \frac{1}{\sqrt 2} \left( v_D + h_D (x) \right) 
\end{equation}
with the VEVs $v$ and $v_D$ fixed by minimisation of the potential
%
%
%
to be
\begin{equation}
\label{vevs}
v^2 = \frac{\mu_\Phi^2 - \frac{1}{2}\frac{\lambda_{\Phi \chi}}{\lambda_\chi} \mu_\chi^2}
{\lambda_\Phi - \frac{1}{4}\frac{\lambda_{\Phi\chi}^2}{\lambda_\chi}}
\;\;\; , \;\;\;
v_D^2 = \frac{\mu_\chi^2 - \frac{1}{2}\frac{\lambda_{\Phi \chi}}{\lambda_\Phi} \mu_\Phi^2}
{\lambda_\chi - \frac{1}{4}\frac{\lambda_{\Phi\chi}^2}{\lambda_\Phi}} \; .
\end{equation}
In terms of the shifted fields $h$ and $h_D$, the scalar potential $V_{\rm scalar}$ can then be decomposed as
$$
V_{\rm scalar} = V_0 + V_1 + V_2 + V_3 + V_4
$$
with
\begin{eqnarray}
V_0 & = & \frac{1}{4} \left( \lambda_\Phi v^4 + \lambda_\chi v_D^4 + \lambda_{\Phi\chi} v^2 v_D^2
- 2 \mu_\Phi^2 v^2 - 2 \mu_\chi^2 v_D^2  \right) \; ,\\
V_1 & = & \frac{1}{2} v \left( 2 \lambda_\Phi v^2 + \lambda_{\Phi \chi} v_D^2 - 2 \mu_\Phi^2 \right) h 
+ \frac{1}{2} v_D \left( 2 \lambda_\chi v_D^2 + \lambda_{\Phi \chi} v^2 - 2 \mu_\chi^2 \right) h_D \; ,\\
\label{masssq}
V_2 & = & \left( \frac{3}{2} \lambda_\Phi v^2 + \frac{1}{4} \lambda_{\Phi \chi} v_D^2 - \frac{1}{2} \mu^2_\Phi \right) h^2
+ \left( \frac{3}{2} \lambda_\chi v_D^2 + \frac{1}{4} \lambda_{\Phi\chi} v^2 - \frac{1}{2} \mu^2_\chi \right) h_D^2 
+ \lambda_{\Phi\chi} v v_D h h_D \; , \nonumber \\
& \equiv & \frac{1}{2} \left( h \; \; h_D \right) \cdot M_S^2  \cdot 
\left( 
\begin{tabular}{c}
$h$\\
$h_D$
\end{tabular}
\right) \; , \\
V_3 & = &  \lambda_\Phi v h^3 + \lambda_\chi v_D h_D^3 + 
\frac{1}{2} \lambda_{\Phi \chi} \left( v_D h_D h^2 + v h h_D^2 \right) \; ,\\
V_4 & = & \frac{1}{4}\lambda_\Phi h^4 + \frac{1}{4} \lambda_\chi h_D^4 + \frac{1}{4}\lambda_{\Phi\chi} h^2 h_D^2 \; .
\end{eqnarray}
Here $V_0$ is a cosmological constant and will be discarded from now on; the tadpole term $V_1$ vanishes with 
$v$ and $v_D$ given by Eq.~(\ref{vevs}); $V_2$ is quadratic in the fields $h$ and $h_D$, and we have to diagonalize 
the mass matrix $M^2_S$ in Eq.~(\ref{masssq}) to get the physical Higgs fields $h_1$ and $h_2$ (see next section); 
and $V_3$ and $V_4$ are the trilinear and quadrilinear self couplings among the two Higgs fields. 
Since $\chi$ is a SM singlet, the $W$ and $Z$ bosons acquire their masses through the SM Higgs doublet
VEV $v$ entirely which implies $v = 246$ GeV.

\section{Mixing Effects}

\subsection{Higgs Mass Eigenstates and Their Self Interactions}

The mass matrix $M_S^2$ in Eq.~(\ref{masssq}) for the scalar bosons is
\begin{eqnarray}
M_S^2 & =& \left(
\begin{tabular}{cc}
$m_{11}^2$ & $m_{12}^2$ \\ 
$m_{21}^2$ & $m_{22}^2$  
\end{tabular}
\right) \; ,\nonumber\\
& = & \left(
\begin{tabular}{cc}
$2 \lambda_\Phi v^2 $ & $\lambda_{\Phi\chi} v v_D$ \\ 
$\lambda_{\Phi\chi} v v_D$ & $2 \lambda_\chi v_D^2$  
\end{tabular}
\right) \; .
\end{eqnarray}
Its eigenvalues are
\begin{equation}
m_{1,2}^2 =  \frac{1}{2} \left[ 
{\rm Tr} M_S^2 \pm \sqrt{\left( {\rm Tr} M_S^2 \right)^2 - 4 \, {\rm Det} M_S^2 } \right] \; .
\end{equation}
The physical Higgs $(h_1, h_2)$ are related to the original $(h, h_D)$ as
\begin{equation}
\left(
\begin{tabular}{c}
$h_1$ \\ 
$h_2$
\end{tabular}
\right)
=
\left(
\begin{tabular}{cc}
$\cos \alpha$ & $\sin \alpha$ \\ 
$-\sin \alpha$ & $\cos \alpha$
\end{tabular}
\right)
\left(
\begin{tabular}{c}
$h$ \\ 
$h_D$
\end{tabular}
\right) \;\; ,
\end{equation}
with the mixing angle 
\begin{equation}
\sin 2 \alpha = \frac{2 m^2_{12}}{m_1^2 - m_2^2} \; .
\end{equation}
We will identify the heavier Higgs $h_1$ with mass $m_1= 126$ GeV
as the new boson observed at the LHC \cite{ATLAS,CMS}, while the lighter one 
$h_2$ has been escaped detection thus far. The SM Higgs couplings with the SM
fermions and gauge bosons are thus modified by a factor of $\cos \alpha$.

In terms of the physical Higgs fields $h_1$ and $h_2$, the cubic term $V_3$ is given by
\begin{equation}
V_3 = \frac{1}{3\\!} \lambda_3^{(1)} h_1^3 + \frac{1}{3\\!} \lambda_3^{(2)} h_2^3 
+ \frac{1}{2} \lambda_3^{(3)} h_1 h_2^2  
+ \frac{1}{2} \lambda_3^{(4)} h_2 h_1^2  
\end{equation}
with the trilinear couplings
\begin{eqnarray}
\lambda_3^{(1)} & = & 3 \left[ 2 v \lambda_\Phi \cos^3 \alpha  + 2 v_D \lambda_\chi \sin^3 \alpha 
+ \frac{1}{2} \lambda_{\Phi\chi} \sin 2 \alpha \left( v \sin \alpha + v_D \cos \alpha \right) \right] \; ,\\
\lambda_3^{(2)} & = & 3 \left[ - 2 v \lambda_\Phi \sin^3 \alpha  + 2 v_D \lambda_\chi \cos^3 \alpha 
+ \frac{1}{2} \lambda_{\Phi\chi} \sin 2 \alpha \left( v_D \sin \alpha - v \cos \alpha \right) \right] \; , \\
\label{couplingh2hdhd}
\lambda_3^{(3)} & = & \frac{1}{4} \left[ 24 v \lambda_\Phi \sin^2 \alpha \cos \alpha 
+ 24 v_D \lambda_\chi \cos^2 \alpha \sin \alpha \right. \nonumber \\
&& \;\;\; \left. + \lambda_{\Phi\chi} \left( v \cos \alpha + v_D \sin \alpha + 3 v \cos 3 \alpha - 3 v_D \sin 3 \alpha \right)
\right] \; , \\
\lambda_3^{(4)} & = & \frac{1}{4} \left[ -24 v \lambda_\Phi \sin \alpha \cos^2 \alpha 
+ 24 v_D \lambda_\chi \cos \alpha \sin^2 \alpha \right. \nonumber \\
&& \;\;\; \left. + \lambda_{\Phi\chi} \left( - v \sin \alpha + v_D \cos \alpha + 3 v \sin 3 \alpha + 3 v_D \cos 3 \alpha \right)
\right] \; ,
\end{eqnarray}
and the quartic term $V_4$ is given by
\begin{equation}
V_4 =  \frac{1}{4\\!} \lambda_4^{(1)} h_1^4 + \frac{1}{4\\!} \lambda_4^{(2)} h_2^4 
+ \frac{1}{3\\!} \lambda_4^{(3)} h_1 h_2^3  
+ \frac{1}{3\\!} \lambda_4^{(4)} h_2 h_1^3
+ \frac{1}{2\\! \cdot 2\\!} \lambda_4^{(5)} h_1^2 h_2^2
\end{equation}
with the quadrilinear couplings
\begin{eqnarray}
\lambda_4^{(1)} & = & 6 \left( \lambda_\Phi \cos^4\alpha + \lambda_\chi \sin^4 \alpha + 
\frac{1}{4} \lambda_{\Phi\chi} \sin^2 2 \alpha \right) \; ,\\
\lambda_4^{(2)} & = & 6 \left( \lambda_\Phi \sin^4\alpha + \lambda_\chi \cos^4 \alpha + 
\frac{1}{4} \lambda_{\Phi\chi} \sin^2 2 \alpha \right) \; , \\
\lambda_4^{(3)} & = & -\frac{3}{2} \sin 2 \alpha \left( -2 \lambda_\chi \cos^2 \alpha + 2 \lambda_\Phi \sin^2 \alpha 
+ \lambda_{\Phi\chi} \cos 2 \alpha \right)  \; , \\
\lambda_4^{(4)} & = & +\frac{3}{2} \sin 2 \alpha \left( 2 \lambda_\chi \sin^2 \alpha - 2 \lambda_\Phi \cos^2 \alpha 
+ \lambda_{\Phi\chi} \cos 2 \alpha \right) \; , \\
\lambda_4^{(5)} & = & \frac{1}{4} \left[ 3 \left( \lambda_\Phi + \lambda_\chi \right) + \lambda_{\Phi\chi} 
- 3 \left( \lambda_\Phi + \lambda_\chi - \lambda_{\Phi\chi} \right) \cos 4 \alpha \right] \; .
\end{eqnarray}
%


\subsection{Kinetic and Mass Mixing of the neutral gauge bosons} 

In additional to the mass mixing of the three neutral gauge bosons arise from 
the spontaneously electroweak symmetry breaking given by
\begin{equation}
{\cal L}_{\rm m} = \frac{1}{2} \left( C^\mu \; B^\mu \; W^{3 \mu} \right) M^2 
\left(
\begin{tabular}{c}
$C_\mu$ \\
$B_\mu$ \\
$W^3_{\mu}$
\end{tabular}
\right)
\end{equation}
with the following mass matrix
\begin{equation}
M^2 = \left(
\begin{tabular}{ccc}
$g^2_D v^2_D$ & 0 & 0\\
0 & $\frac{1}{4}g'^2 v^2$ & $-\frac{1}{4} g g' v^2$\\
0 & $-\frac{1}{4} g g' v^2$ & $\frac{1}{4}g^2 v^2$
\end{tabular}
\right) \; ,
\end{equation}
we also have 
the kinetic mixing between the two $U(1)$s from the last term in Eq.~(\ref{Lgauge}).
Both the kinetic and mass mixings can be diagonalized simultaneously by the following 
mixed transformation \cite{Feldman-Liu-Nath}
\begin{equation}
\left(
\begin{tabular}{c}
$C_\mu$ \\
$B_\mu$ \\
$W^3_\mu$
\end{tabular}
\right) =
K \cdot O
\left(
\begin{tabular}{c}
$A'_\mu$ \\
$Z_\mu$ \\
$A_\mu$
\end{tabular}
\right)
\end{equation}
where $A'_\mu$, $Z_\mu$ and $A_\mu$ are the physical dark photon, $Z$ boson and the photon respectively.
Here $K$ is a general linear transformation that diagonalizes the kinetic mixing
\begin{equation}
K = \left(
\begin{tabular}{ccc}
$\beta$ & 0 & 0 \\
$- \epsilon \beta$ &  1  & 0 \\
0 & 0 & 1
\end{tabular}
\right) \; , 
\end{equation}
where $\beta = 1/(1-\epsilon^2)^{1/2} \, (\epsilon \leq 0)$, 
and $O$ is a $3 \times 3$ orthogonal matrix which can be parametrized as
\begin{equation}
 O = \left(
\begin{tabular}{ccc}
$\cos\psi \cos \phi - \sin \theta \sin \phi \sin \psi$ \; & \; $\sin \psi \cos \phi + \sin \theta \sin \phi \cos \psi$ 
\; & \; $- \cos \theta \sin \phi$ \\
$\cos\psi \sin \phi + \sin \theta \cos \phi \sin \psi$ \; & \; $\sin \psi \sin \phi - \sin \theta \cos \phi \cos \psi$ 
\; & \; $ \cos \theta \cos \phi$ \\
$- \cos \theta \sin \psi$ \; & \; $\cos \theta \cos \psi$ 
\; & \; $\sin \theta $
\end{tabular}
\right)
\end{equation}
with the mixing angles given by \cite{Feldman-Liu-Nath}
\begin{equation}
\tan\theta = \frac{g'}{g},\quad  \tan\phi = -\epsilon \beta,\quad   \tan\psi = \pm{} \frac{\tan\phi \cos\theta}{\tan\theta} \left[\frac{1-M^2_Z/M^2_W}{1-M^2_Z/g_D^2v_D^2} + \tan^2\theta \right]  \; .
\end{equation}
After the $K$ transformation, the gauge boson mass matrix is
\begin{equation}
{\tilde M}^2 = K^T M^2 K = 
\left(
\begin{tabular}{ccc}
$ \beta^2\left( g^2_D v^2_D + \frac{1}{4} \epsilon^2 g'^2 v^2 \right)  $ & \, $- \frac{1}{4} \epsilon \beta g'^2 v^2 $ \, 
& $\frac{1}{4} \epsilon \beta g g' v^2 $\\
$- \frac{1}{4} \epsilon \beta g'^2 v^2 $ & $\frac{1}{4}g'^2 v^2$ & $-\frac{1}{4} g g' v^2$\\
$\frac{1}{4} \epsilon \beta g g' v^2 $ & $-\frac{1}{4} g g' v^2$ & $\frac{1}{4}g^2 v^2$
\end{tabular}
\right) \; .
\end{equation}
The $O$ matrix diagonalizes this ${\tilde M}^2$ matrix 
\begin{equation}
M^2_{\rm Diag} = O^T {\tilde M}^2 O = 
\left(
\begin{tabular}{ccc}
$M^2_{\gamma_D}$ & 0 & 0 \\
0 & $M^2_Z$ & 0 \\
0 & 0 & $M^2_\gamma$
\end{tabular}
\right) 
\end{equation}
with the following eigenvalues (assuming $M_{\gamma_D} \leq M_{Z}$
\footnote{For the case of $M_{\gamma_D} > M_{Z}$, we 
will have $M^2_{\gamma_D,Z}  =  (q \pm p)/2$ as was studied by the authors 
in \cite{Feldman-Liu-Nath}.})
\begin{equation}
M^2_\gamma = 0 \; \; , \quad \quad M^2_{Z,\gamma_D}  =  (q \pm p)/2 
\end{equation}
where
\begin{eqnarray}
p & = & \sqrt{q^2  - g^2_D v^2_D v^2 (g^2 + g'^2) \beta^2} \; , \\
q & = & g_D^2 v_D^2 \beta^2 + \frac{1}{4}\left( g^2 + g'^2 \beta^2 \right) v^2 \; .
\end{eqnarray}
For small kinetic parameter mixing $\epsilon$, the $Z$ and $\gamma_D$ masses  can be approximated by 
$M_Z \approx \sqrt{(g^2 + g'^2)} v /2$ and $M_{\gamma_D} \approx g_D v_D $. 

For couplings of these physical neutral gauge bosons with the SM fermions, we refer the readers to
Ref.~\cite{Feldman-Liu-Nath}.


\section{Non-standard Decays of $h_1$}

The global fits \cite{globalfits-Kingman,globalfits-Strumia,globalfits-Ellis,globalfits-Gunion}
for the signal strengths of the various SM Higgs decay channels from the LHC data 
imply the total width of the SM Higgs is about 4.03 MeV and
the non-standard width for the SM Higgs can be at most 1.2 MeV; in other words the non-standard branching ratio 
for the SM Higgs must be less than 22\%. One can use this result to constrain the parameter space of the 
model. 

We will compute the following non-standard processes 
$h_1 \to \gamma_D \gamma_D$, $h_1 \to h_2 h_2$, $h_1 \to h_2 h_2^* \to h_2 \gamma_D \gamma_D$ and
$h_1 \to h_2 h_2 h_2$. Each of the $h_2$ in the final state of these processes will decay into two dark photons 
and each dark photon will give rise to two leptons through its mixing with the photon 
\footnote{We note that $h_2$ can decay to SM particles as well through its mixing with $h_1$ and hence 
they are suppressed. We take the branching ratio of $h_2 \to \gamma_D \gamma_D$ to be 100\%. See discussion after
Eq.~(\ref{h2tohDhD}).}. 
These non-standard processes will provide multiple leptons in the final state of the standard model 
Higgs decay \cite{Zupan-1}. 
The contribution to the heavier Higgs width from these non-standard processes is 
\footnote{The four lepton modes from the first term 
$h_1 \to \gamma_D \gamma_D$ followed by $\gamma_D \to l \bar l \, (l = e,\mu)$ were studied in details
in \cite{Gopalakrishna:2008dv}.}
\begin{equation}
\Gamma^{NS}_{h_1} = \sin^2 \alpha \hat \Gamma(h_1 \to \gamma_D \gamma_D) 
+ \Gamma(h_1 \to h_2 h_2)
+ \Gamma(h_1 \to h_2 \gamma_D \gamma_D)
+ \Gamma(h_1 \to h_2 h_2 h_2) + \cdots
\end{equation}
Thus the total width of the heavier Higgs $h_1$ is modified as 
\begin{equation}
\Gamma_{h_1}  =  \cos^2 \alpha \hat \Gamma_{h} + \Gamma^{NS}_{h_1} \; ,
\end{equation}
where $\hat \Gamma_{h}$ is the width of the SM Higgs $h$, which has a theoretical value of 4.03 MeV. 
The branching ratio for the non-standard modes of the heavier Higgs decay is
\begin{equation}
\label{BrNSh1}
B^{NS}_{h_1} = \frac{\Gamma^{NS}_{h_1}}{\Gamma_{h_1} } \; ,
\end{equation}
which should be constrained to be less than 22\% or so.
The partial decay width for the two body decays are given by
\begin{equation}
\label{h1togDgD}
\hat \Gamma(h_1 \to \gamma_D \gamma_D) = 
\frac{g_D^2 m^2_{\gamma_D}}{8 \pi m_1} 
\left( 1 - \frac{4 m^2_{\gamma_D}}{m^2_1} \right)^{\frac{1}{2}} 
\left( 3 - 
\frac{m_1^2}{m^2_{\gamma_D}} + \frac{m_1^4}{4 m^4_{\gamma_D}} 
\right) \; ,
\end{equation}
and 
\begin{equation}
\label{h1tohDhD}
\Gamma \left( h_1 \to h_2 h_2 \right) = 
\frac{ \left( \lambda^{(3)}_3 \right)^2 }{32 \pi m_1} 
\left( 1 - \frac{4 m_2^2}{m_1^2} \right)^{\frac{1}{2}} \; .
\end{equation}
For the three body decay $h_1 \to h_2 h_2 h_2$, we obtain
\begin{equation}
\label{h1toh2h2h2}
\Gamma \left( h_1 \to h_2 h_2 h_2 \right) = \int_{x_1^{min}}^{x_1^{max}} dx_1 
\int_{x_2^{min}}^{x_2^{max}} dx_2
\frac{d \Gamma \left( h_1 \to h_2 h_2 h_2 \right)}{dx_1 dx_2}
\end{equation}
with the following differential decay rate
\begin{equation}
\frac{d \Gamma \left( h_1 \to h_2 h_2 h_2 \right)}{dx_1 dx_2} = 
\frac{m_1}{1536 \pi^3} \vert {\cal M} \vert^2
\end{equation}
where the matrix element is given by
\begin{equation}
\label{meh1to3h2}
 {\cal M}  = \lambda_4^{(3)} + \frac{1}{m_1^2}
 \left(
  \lambda_3^{(2)} \lambda_3^{(3)} \sum_{i=1,2,3} \left( 1 - x_i \right)^{-1} 
 + \lambda_3^{(3)} \lambda_3^{(4)}  \sum_{i=1,2,3} \left(  \mu - x_i \right)^{-1}
 \right) \; ,
 \end{equation} 
with $\mu = m_2^2/m_1^2$ and $x_1 + x_2 + x_3 = 2$. The range of integration for $x_1$ and $x_2$ 
is confined by
\begin{equation}
2 \sqrt{\mu}  <  x_1 < 1 - 3 \mu \; ,
\end{equation}
\begin{equation}
x_2 {}^{<}_{>} \frac{1}{2} \left( 1 + \mu -x_1 \right)^{-1} \left[ \left( 2 - x_1 \right) \left( 1 + \mu - x_1 \right) 
\pm \left( x_1^2 - 4 \mu \right)^{1/2} \lambda^{1/2} \left( 1 + \mu - x_1, \mu, \mu \right) \right]
\end{equation}
where $\lambda (a,b,c) = a^2 + b^2 + c^2 - 2 (ab + bc + ca)$. 

The matrix element for the three body process $h_1 \to h_2 \gamma_D \gamma_D$ is rather
long, we will not present the expression here but it is included entirely in our numerical work.

Now the dark Higgs $h_2$ decays into $\gamma_D \gamma_D$ and SM particles with coefficients 
$\cos^2 \alpha$ and $\sin^2 \alpha$ respectively, so its branching fraction into 
$\gamma_D \gamma_D$ is given by
\begin{equation}
\label{h2tohDhD}
B(h_2 \to \gamma_D \gamma_D) = \frac{\cos^2 \alpha \hat \Gamma (h_2 \to \gamma_D \gamma_D)} 
{\cos^2 \alpha \hat \Gamma (h_2 \to \gamma_D \gamma_D) + \sin^2 \alpha \hat \Gamma^{\rm SM}_{h_2}} \; ,
\end{equation}
where $\hat \Gamma (h_2 \to \gamma_D \gamma_D)$ can be obtained from Eq.~(\ref{h1togDgD}) 
with the following substitution $m_1 \to m_2$, and
$\hat \Gamma^{\rm SM}_{h_2}$ is the partial decay width of $h_2$ into SM particles. 
Since $\hat \Gamma^{\rm SM}_{h_2}$ are suppressed by a factor of $\sin^2 \alpha$, the above
branching fraction is close to unity.

\begin{figure}[t!]
\centering
\includegraphics[width=4in]{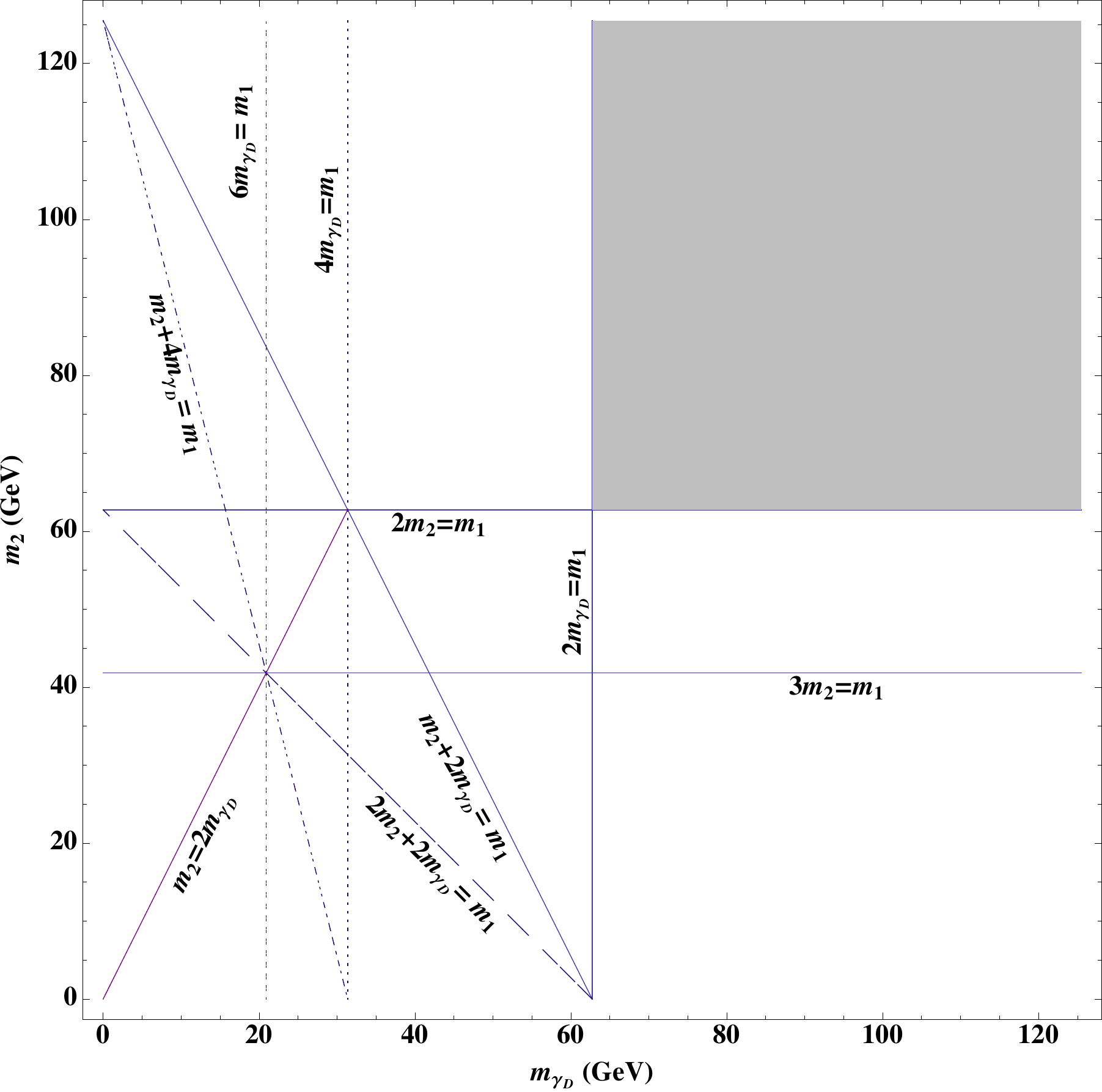}
\caption{\small \label{Regions}
The kinematical regions in the $(m_{\gamma_D}, m_2)$ plane for the non-standard decays of the heavier Higgs $h_1$,
identified as the 126 GeV boson observed at LHC.
See the last paragraph of section IV  for illustrations.
}
\end{figure}

In Fig.~\ref{Regions}, the various regions of kinematics in the $(m_{\gamma_D},m_2)$ plane
that exhibits the very rich Higgs phenomenology in this model are schematically shown. 
The different regions can be described briefly as follows:
\begin{itemize}

\item
Clearly, the two lines $2 m_{\gamma_D} = m_1$ (left of which $h_1 \to \gamma_D \gamma_D$ is open) 
and $2 m_2 = m_1$  (below of which $h_1 \to h_2 h_2$ is open) defines our region of interest (un-shaded). 

\item
Below the line $3 m_2 = m_1$, the 3-body process $h_1 \to h_2 h_2 h_2$ is open too. 

\item
Other lines correspond to 2, 4, or 6 dark photons coming from the decays of $h_2$ in $h_1 \to h_2 h_2$ or
$h_1 \to h_2 h_2 h_2$:
{\it i.e.} to the left of the 5 lines
$2 m_{\gamma_D} + m_2= m_1$, 
$2 m_{\gamma_D} + 2 m_2= m_1$,
$4 m_{\gamma_D} + m_2= m_1$,
$4 m_{\gamma_D} = m_1$ and
$6 m_{\gamma_D} = m_1$
correspond to the openings of the 5 processes
$h_1 \to h_2 \gamma_D \gamma_D$,
$h_1 \to h_2 h_2 \gamma_D \gamma_D$,
$h_1 \to h_2 \gamma_D \gamma_D \gamma_D \gamma_D$,
$h_1 \to \gamma_D \gamma_D \gamma_D \gamma_D$ and
$h_1 \to \gamma_D \gamma_D \gamma_D \gamma_D \gamma_D \gamma_D$ respectively.

\item
Lastly, the special line $m_2 = 2 m_{\gamma_D}$ emanated from the coordinate origin 
separates the $\gamma_D\gamma_D$ pair coming from either a on-shell $h_2$ or off-shell $h_2^*$
for these multi-$\gamma_D$ processes. 

\end{itemize}

Since the dark photon $\gamma_D$ will mix with the photon,
through either kinetic mixing \cite{Holdom} or a gauge invariant Stueckelberg mass term \cite{Stueckelberg}, 
it will communicate with the SM fermions eventually. 
If the dark photon mass is larger than twice the electron or muon mass, theses processes will lead to multileptons 
in the final states of the $h_1$ decay. 
These lepton jets can be distinguished from the QCD jets by imposing cuts on the electromagnetic ratio 
and charge ratio, as proposed in \cite{Zupan-1}. 
Supersymmetric models with or without R-parity can also give rise to multilepton events as
experimentalists had searched for such signals and placed exclusion limits on the masses 
of supersymmetric particles \cite{Chatrchyan:2012mea}.
LHC search for multilepton Higgs decay modes in the dark portal model 
will be discussed later in Sec. 6.


\section{Branching Ratios}

In our numerical work, we will restrict our interest where both the dark photon and dark Higgs 
have masses smaller than 126 GeV. In particular, we will pay special attention to the small mass region
where their masses are in the range of 0.5 to a few GeV. In this range, final states of 
$\tau$ pair and light quarks pairs (pion and kaon pairs) from the dark photon decay are also possible, 
but they are harder to detect at the LHC. 

Limit for invisibly decay of a Higgs boson with mass as low as 1 GeV had been
reported by OPAL \cite{opal} \footnote{We would like to thank W.~Y. Keung bringing us the attention of this experimental paper.}. For a 1 GeV Higgs boson mass, an upper limit for the mixing angle 
of $\vert \alpha \vert \le 3 \times 10^{-2}$ can be extracted from the Fig.~5 in Ref.~\cite{opal}. However
the exclusion curve on the Higgs mass versus mixing angle plot given in \cite{opal} 
was obtained under the assumption that invisible branching ratio of the Higgs boson decay is 100\%. 
Relaxing this assumption would lead to larger mixing angle for a given Higgs mass. In the present case,
the branching ratio in Eq.~(\ref{BrNSh1}) must be less than 20\% or so.

\begin{figure}[t!]
\centering
\includegraphics[width=2.8in]{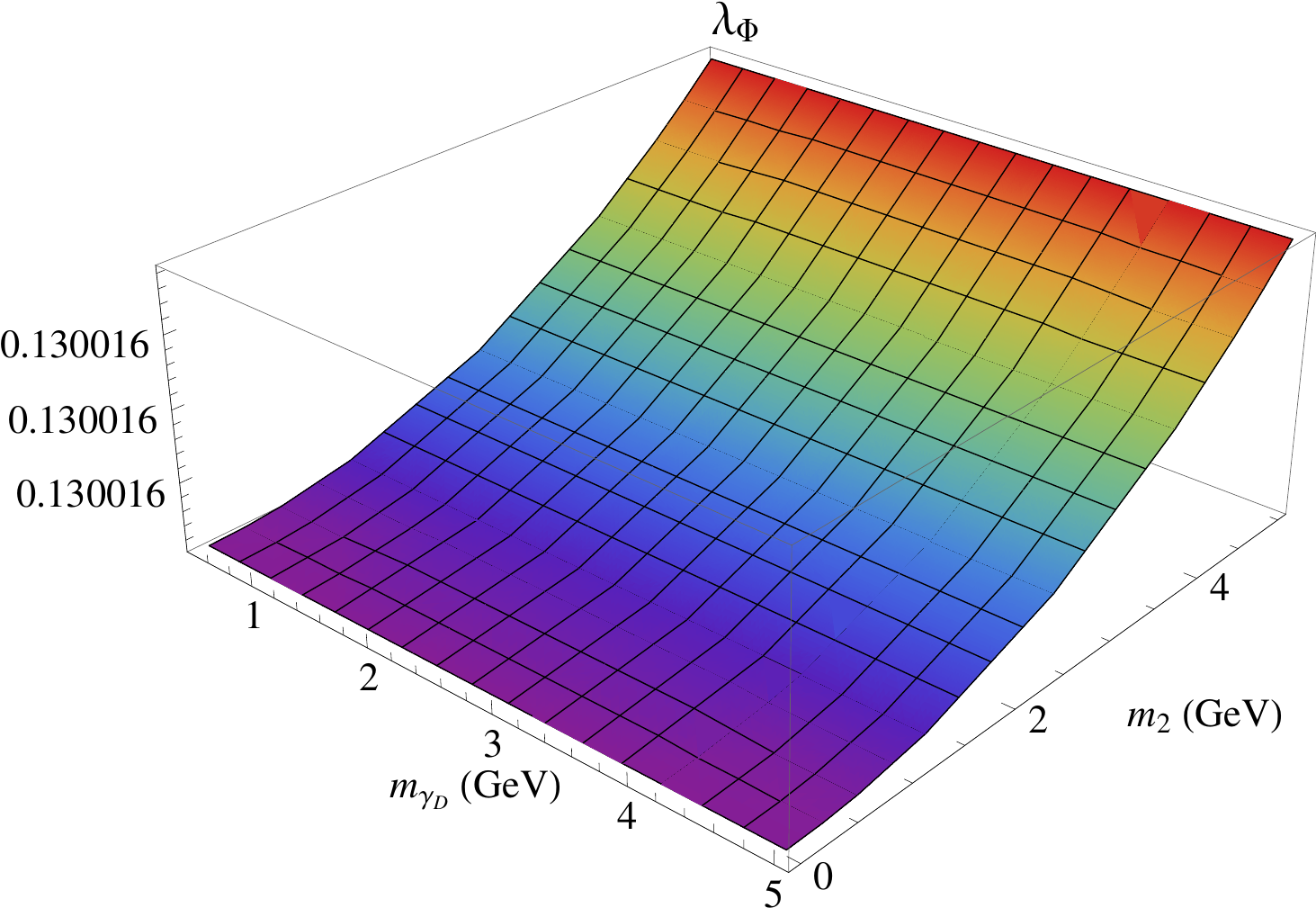}
\includegraphics[width=2.8in]{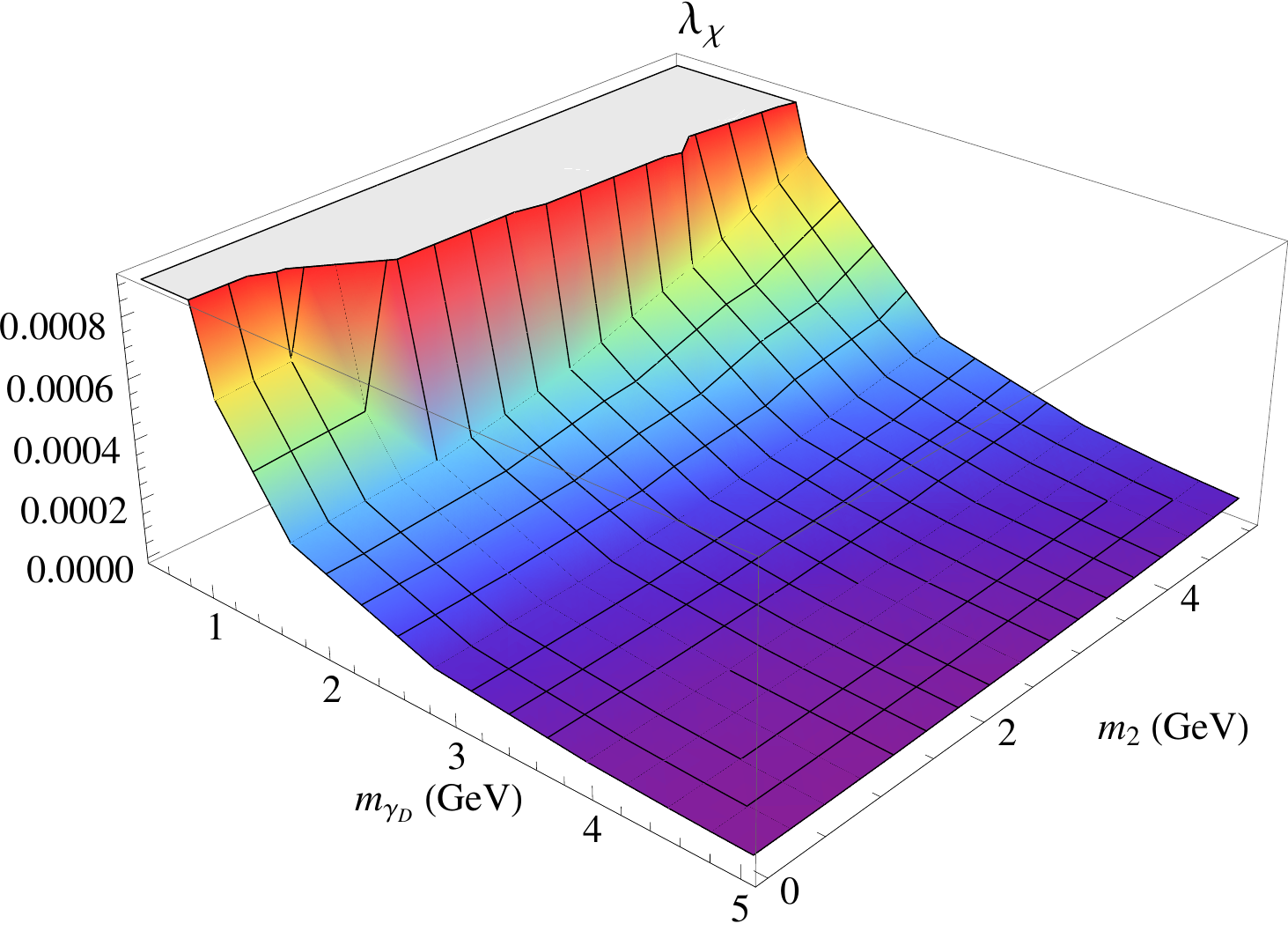}
\includegraphics[width=2.8in]{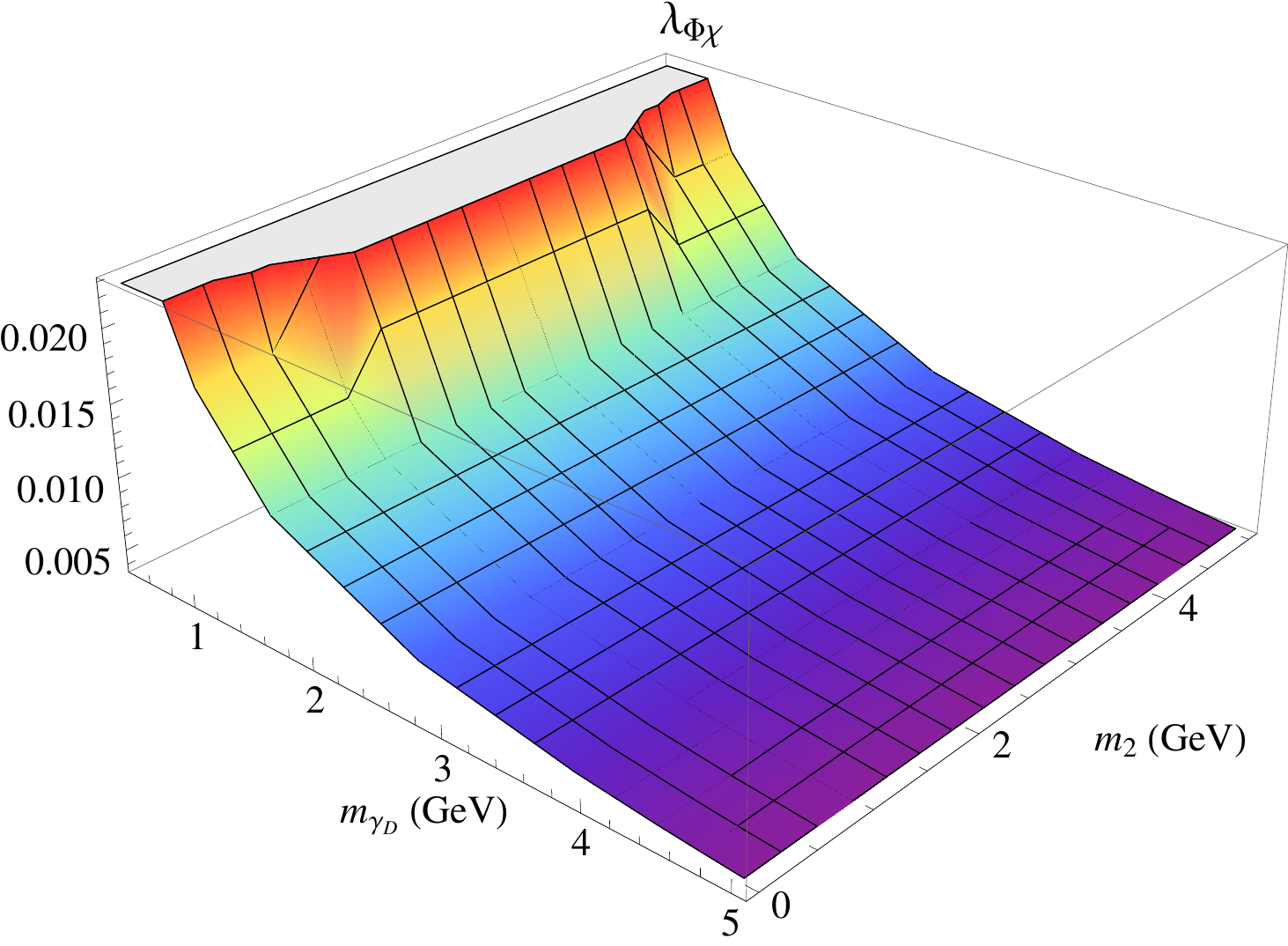}
\includegraphics[width=2.8in]{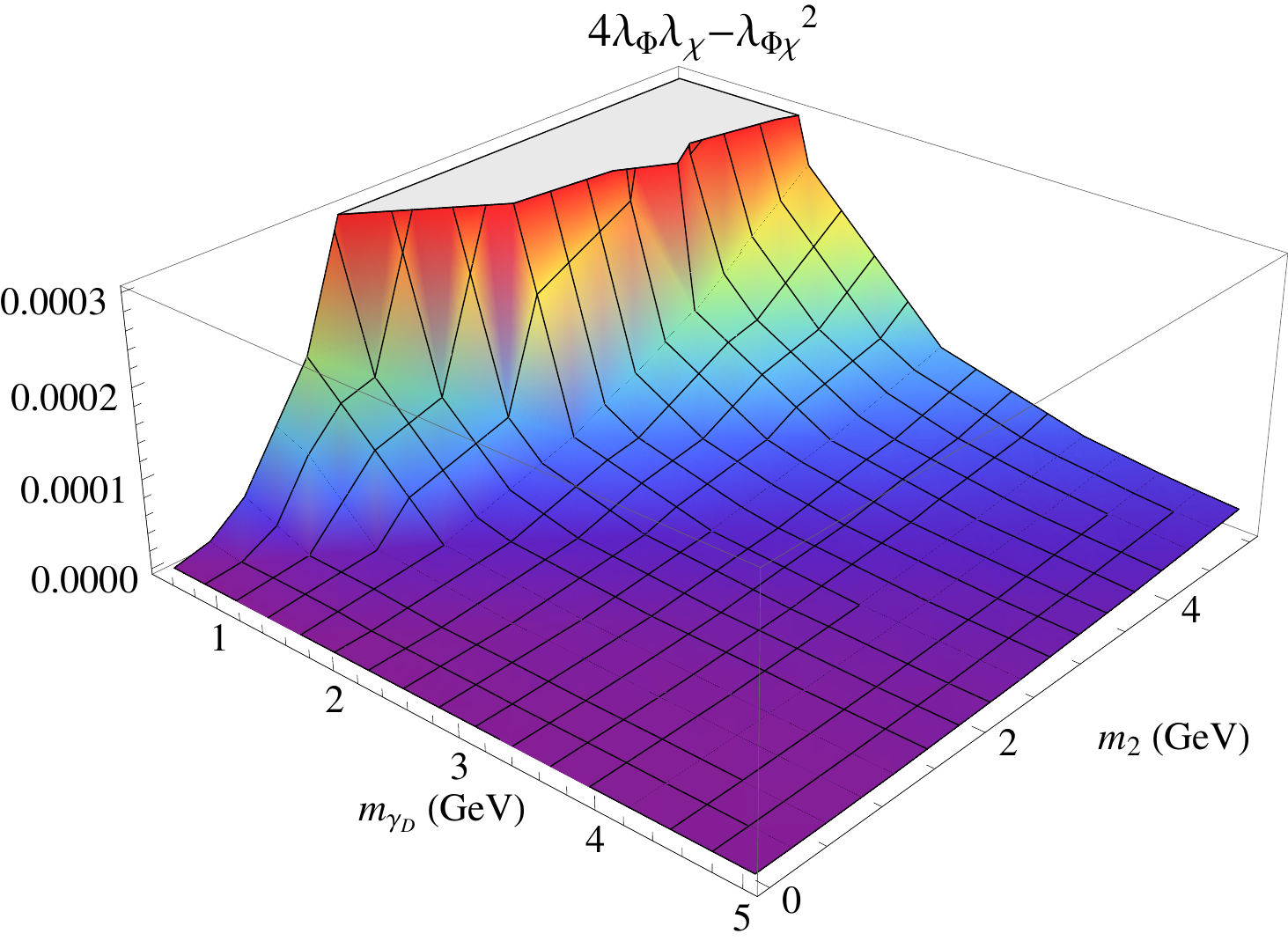}
\caption{\small \label{lambdas}
Fundamental couplings $\lambda_\Phi$, $\lambda_\chi$ and $\lambda_{\Phi\chi}$, and their combination
$(4 \lambda_\Phi \lambda_\chi - \lambda_{\Phi\chi}^2)$ as function of 
$(m_{\gamma_D}, m_2)$ in the small mass region up to 5 GeV with fixed values of $g_D = 0.01$ and $\alpha = 0.03$.
}
\end{figure}
In Fig.~\ref{lambdas},
we plot the fundamental couplings $\lambda_\Phi$, $\lambda_\chi$ and $\lambda_{\Phi\chi}$ that 
entered in the Lagrangian density and their 
combination $(4 \lambda_\Phi \lambda_\chi - \lambda_{\Phi\chi}^2)$ as function of 
$(m_{\gamma_D}, m_2)$ in the small mass region up to 5 GeV with fixed values of $g_D = 0.01$ and $\alpha = 0.03$.
As one can easily see that $\lambda_\Phi$ is not sensitive to these input parameters 
and very close to its SM value of $m_1^2/2v^2 = 0.13$. 
We note the following hierarchy
$\lambda_{\chi} \ll  \lambda_{\Phi \chi} \ll \lambda_{\Phi}$ in this small mass region from the first three plots of this figure.
Moreover, the positiveness of the combination
$(4 \lambda_\Phi \lambda_\chi - \lambda_{\Phi\chi}^2)$ in the last plot of this figure implies the scalar 
potential is bounded from below at tree level.

\begin{figure}[t!]
\centering
\includegraphics[width=2.6in]{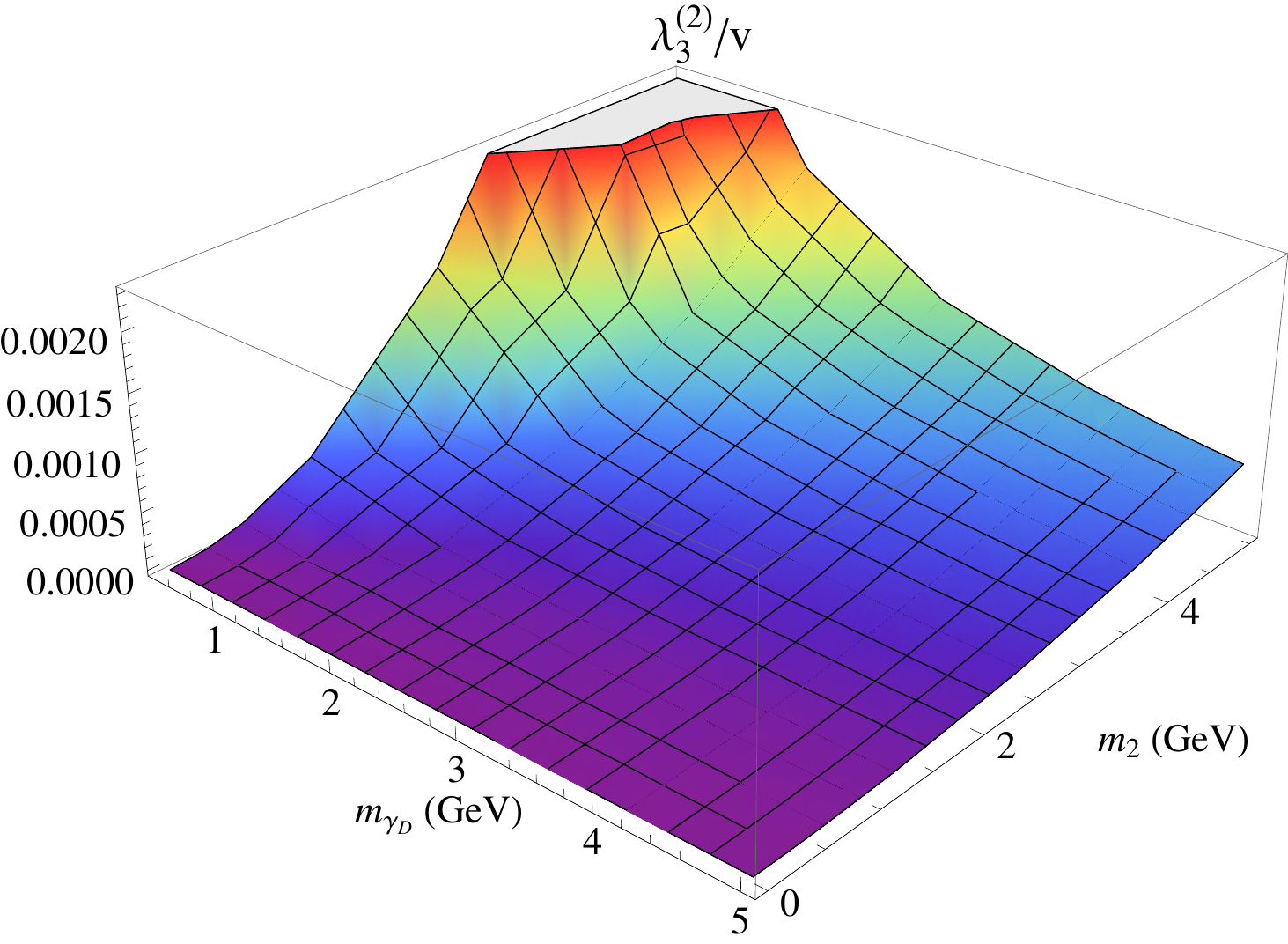}
\includegraphics[width=2.6in]{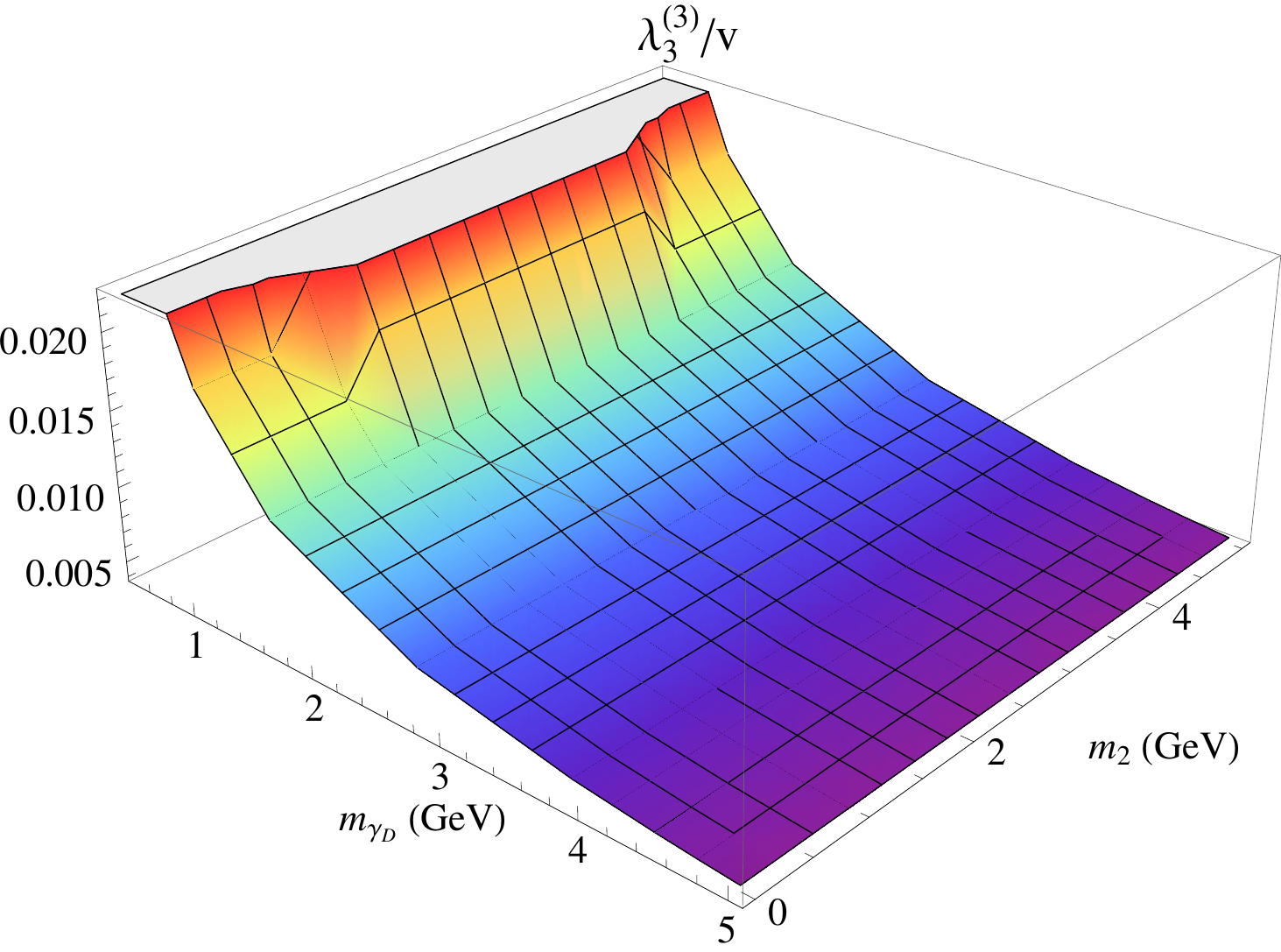}
\includegraphics[width=2.6in]{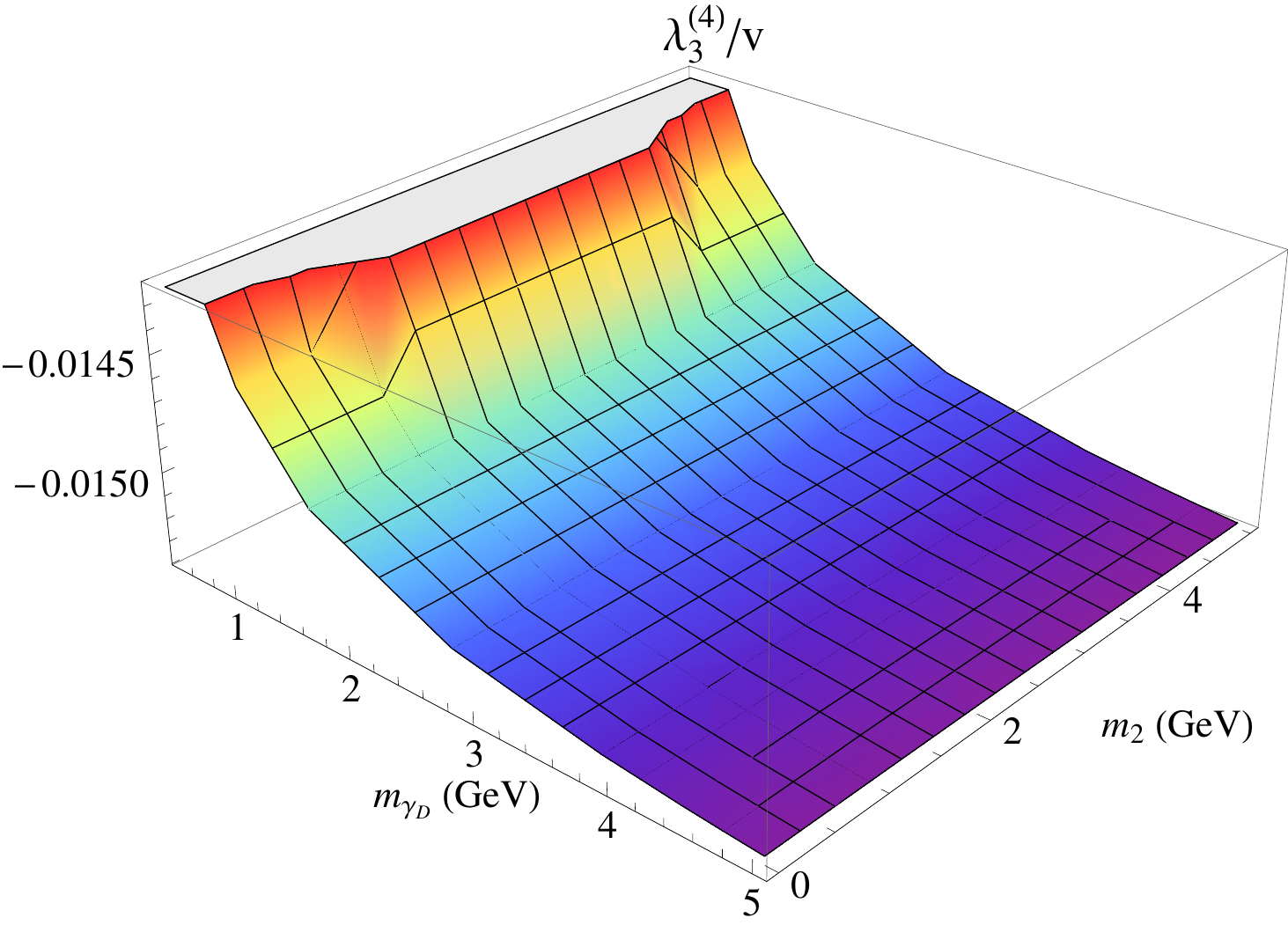}
\includegraphics[width=2.6in]{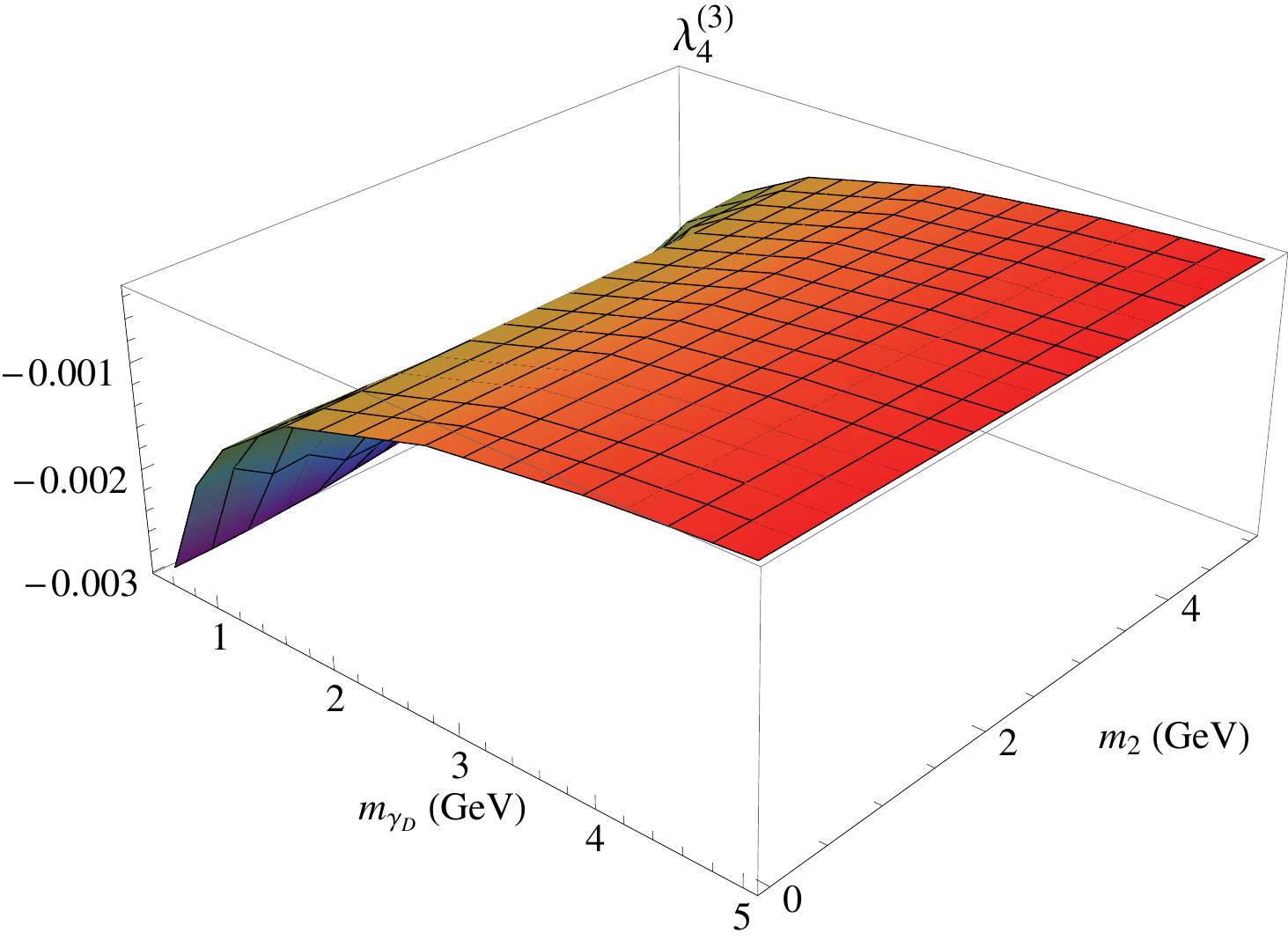}
\caption{\small \label{couplings34}
Couplings $\lambda_3^{(2)}/v$, $\lambda_3^{(3)}/v$, $\lambda_3^{(4)}/v$ and
$\lambda_4^{(3)}$ as function of 
$(m_{\gamma_D}, m_2)$ in the small mass region up to 5 GeV with fixed values of $g_D = 0.01$ and $\alpha = 0.03$.
}
\end{figure}
In Fig.~\ref{couplings34},
we plot the trilinear couplings $\lambda_3^{(2)}/v$, $\lambda_3^{(3)}/v$ and $\lambda_3^{(4)}/v$ 
normalized to the VEV $v$, and the quadrilinear coupling $\lambda_4^{(3)}$ that are relevant to the three body processes 
$h_1 \to h_2 \gamma_D \gamma_D$ and $h_1 \to h_2 h_2 h_2$ as function of 
$(m_{\gamma_D}, m_2)$ in the small mass region up to 5 GeV with fixed values of $g_D = 0.01$ and $\alpha = 0.03$.
We note that the matrix element (Eq.~(\ref{meh1to3h2})) for the three body process $h_1 \to h_2 h_2 h_2$ 
involves one term proportional to the quartic coupling $\lambda_4^{(3)}$ and two other terms proportional to 
the product of cubic couplings $\lambda_3^{(2)} \cdot \lambda_3^{(3)}$ 
and $\lambda_3^{(3)} \cdot \lambda_3^{(4)}$ respectively,
while the matrix element for the three body process $h_1 \to h_2 \gamma_D \gamma_D$ involves 
diagrams proportional to $g_D^2 \sin \alpha \cos \alpha$, $g_D^2 \lambda_3^{(3)} \cos \alpha$, 
$g_D^2  \lambda_3^{(4)} \sin \alpha$ or $g_D^4 \sin \alpha \cos \alpha$. The dark gauge coupling $g_D$ alone
in general is not too severely constrained by experiments \footnote{At the low mass region 
of the dark photon and dark Higgs that we are interested in, the {\it B{\scriptsize A}B{\scriptsize AR}} 
experiment \cite{BaBar} had only obtained the limit for the product $\alpha_D \cdot \epsilon^2$, 
where $\alpha_D=g_D^2/4 \pi$ and $\epsilon$ is the kinetic  mixing parameter in Eq.~(\ref{Lgauge}), 
as a function of the dark Higgs mass or dark photon mass.}. 
On the other hand, since the 126 GeV new boson observed at the LHC behaves very much
SM-like, the mixing angle $\alpha$ is constrained to be quite small.
Thus the three body decay $h_1 \to h_2 \gamma_D \gamma_D$ 
is expected to be more relevant than $h_1 \to h_2 h_2 h_2$. In our analysis, we include both of these three body modes 
and find that the mode $h_1 \to h_2 h_2 h_2$ is indeed negligible.

\begin{figure}[t!]
\centering
\includegraphics[width=2.5in,height=2.5in]{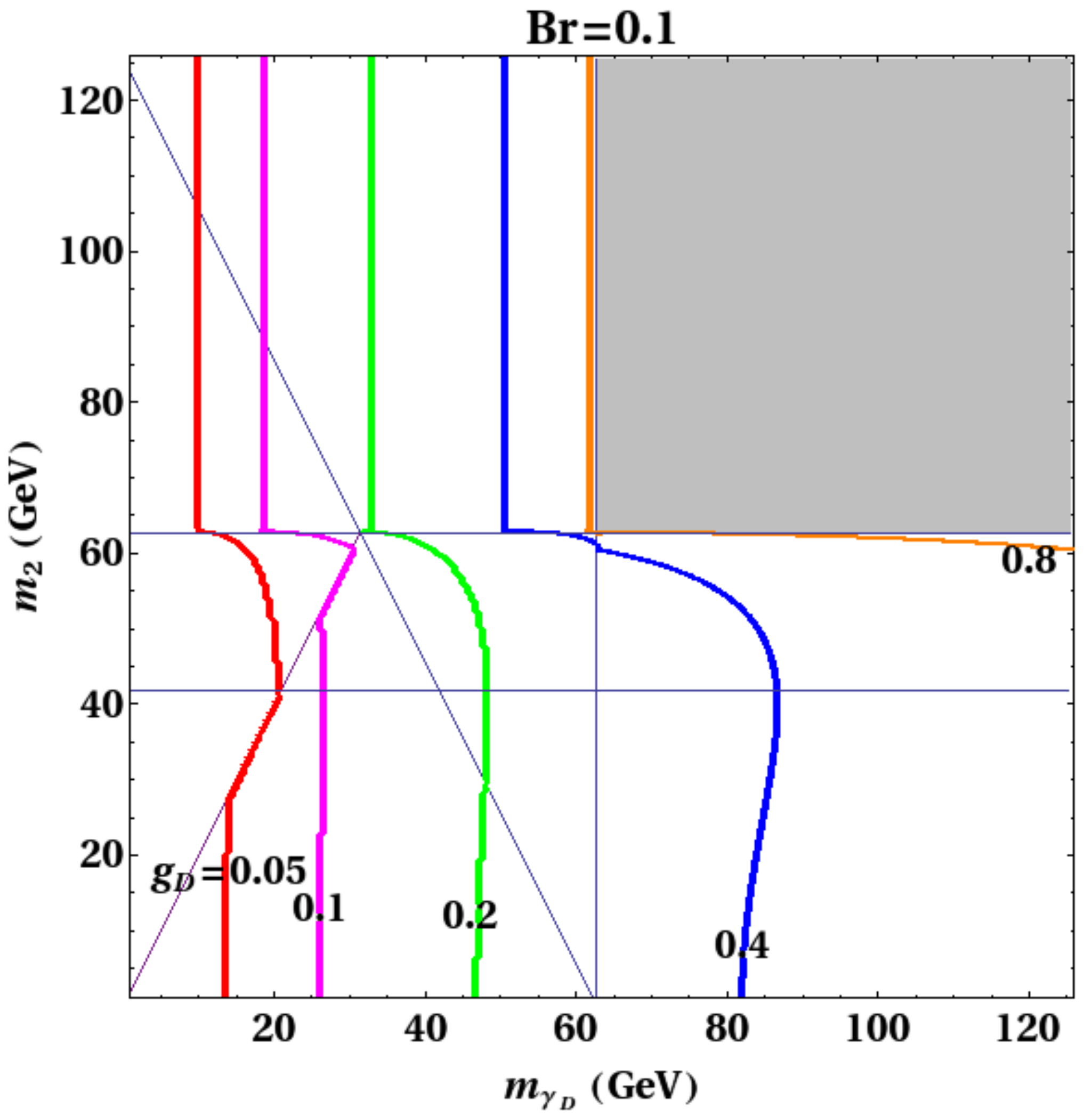}
\includegraphics[width=2.5in,height=2.5in]{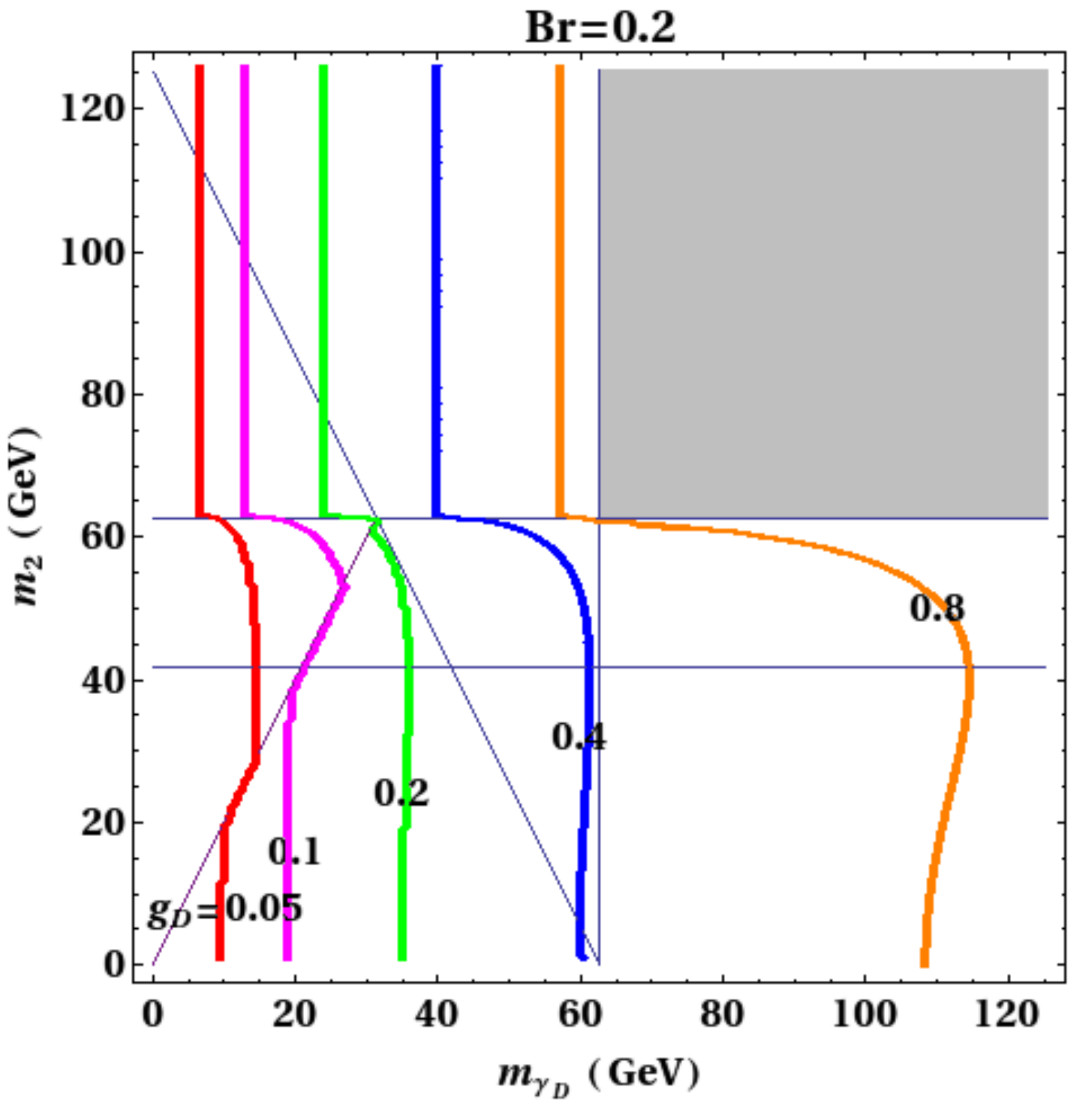}
\caption{\small \label{branching1}
Contour plot of the non-standard branching ratio $B^{NS}_{h_1}$ (Eq.~(\ref{BrNSh1})) = 0.1 (left) and 0.2 (right) 
of the heavier Higgs $h_1$ in the $(m_{\gamma_D},m_2)$ plane up to 126 GeV in both directions 
for $\sin^2\alpha = 0.0009$ and $g_D$ = 0.05, 0.1, 0.2, 0.4 and 0.8.}
\end{figure}

In Fig.~\ref{branching1}, we plot the contour of the non-standard branching ratio $B^{NS}_{h_1}$ (Eq.~(\ref{BrNSh1})) 
= 0.1 (left) and 0.2 (right) of the heavier Higgs $h_1$ in the $(m_{\gamma_D},m_2)$ plane up to 126 GeV  
in both directions with the following parameter input: $\sin^2\alpha = 0.0009$ and $g_D$ = 0.05, 0.1, 0.2, 0.4 and 0.8.
%

\begin{figure}[t!]
\centering
\includegraphics[width=2.5in,height=2.5in]{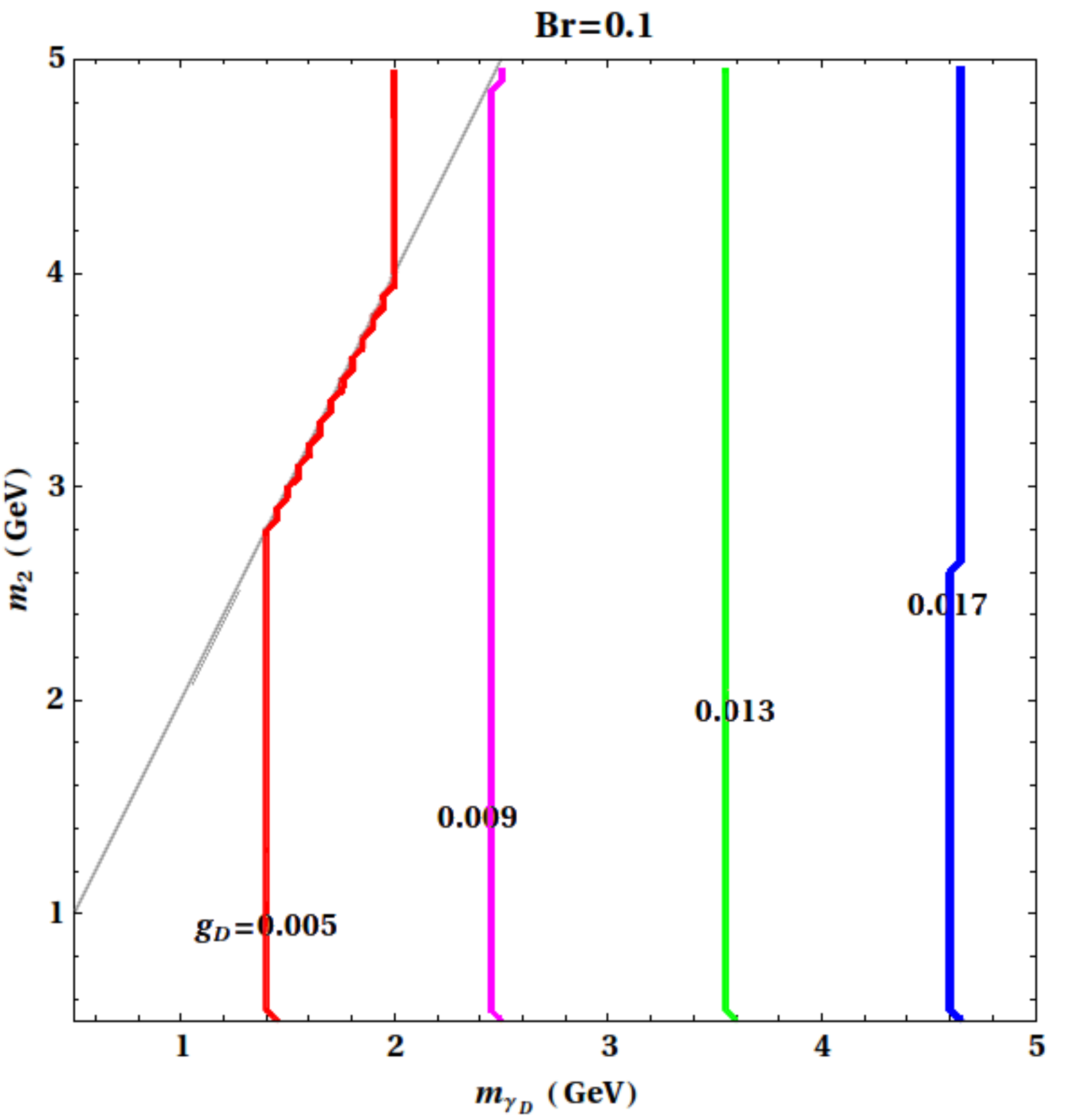}
\includegraphics[width=2.5in,height=2.5in]{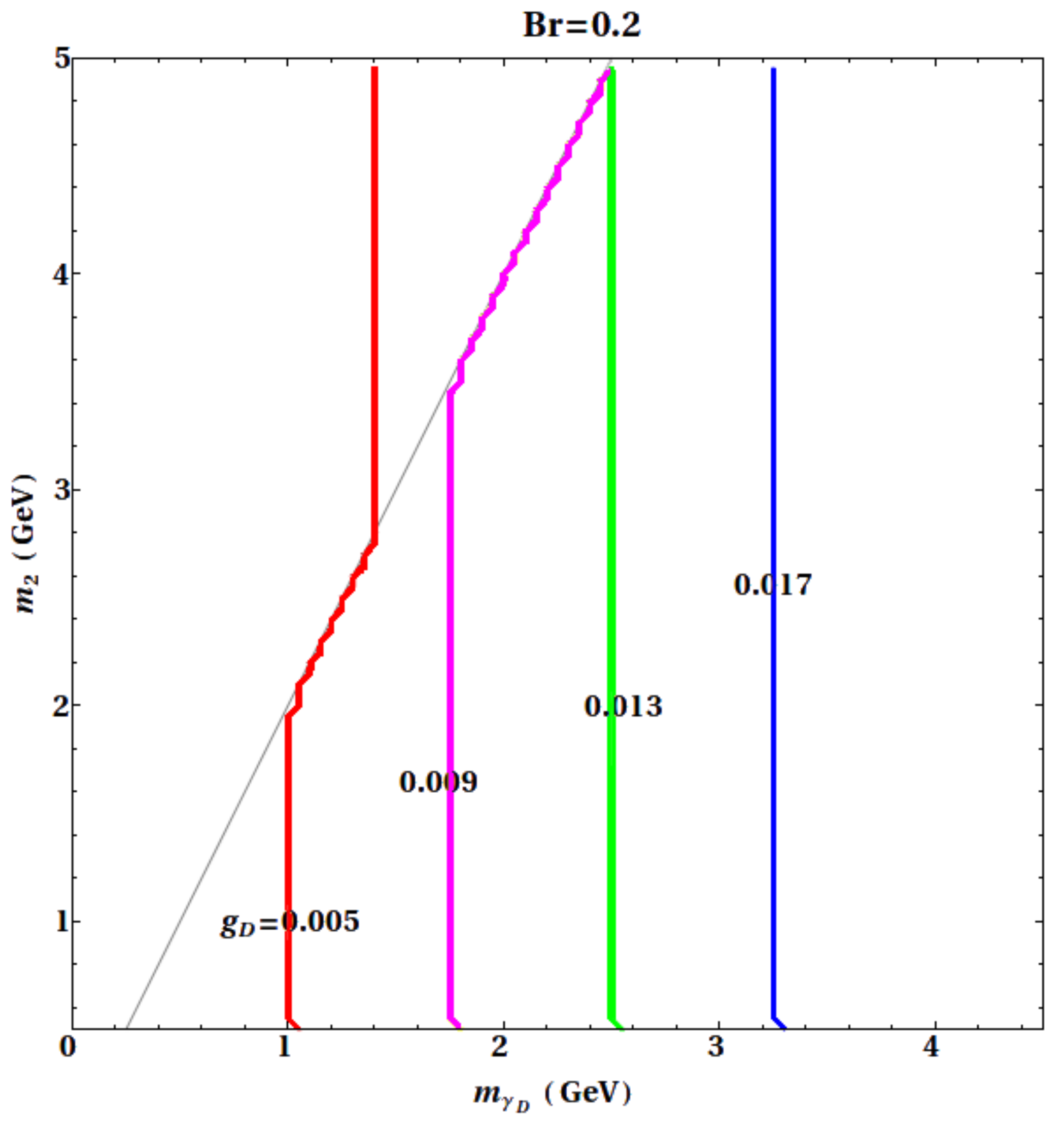}
\caption{\small \label{branching2}
Contour plot of the non-standard branching ratio $B^{NS}_{h_1}$ (Eq.~(\ref{BrNSh1})) = 0.1 (left) and 0.2 (right)
of the heavier Higgs $h_1$ in the small mass 
region of 0.5 to 5 GeV in the $(m_{\gamma_D},m_2)$ plane for $\sin^2\alpha = 0.0009$ and $g_D$ = 0.005, 0.009, 0.013 and 0.017.
}
\end{figure}

In Fig.~\ref{branching2}, we plot the contour of the non-standard branching ratio 
$B^{NS}_{h_1}$ (Eq.~(\ref{BrNSh1})) = 0.1 (left) and 0.2 (right) of the heavier Higgs $h_1$ 
in the $(m_{\gamma_D},m_2)$ plane for the small mass region of 0.5 to 5 GeV in both directions 
for $\sin^2\alpha = 0.0009$ and $g_D$ = 0.005, 0.009, 0.013 and 0.017.
%

\section{Multilepton Jets at the LHC}

\begin{figure}[h!]
\centering
\includegraphics[width=2.4in]{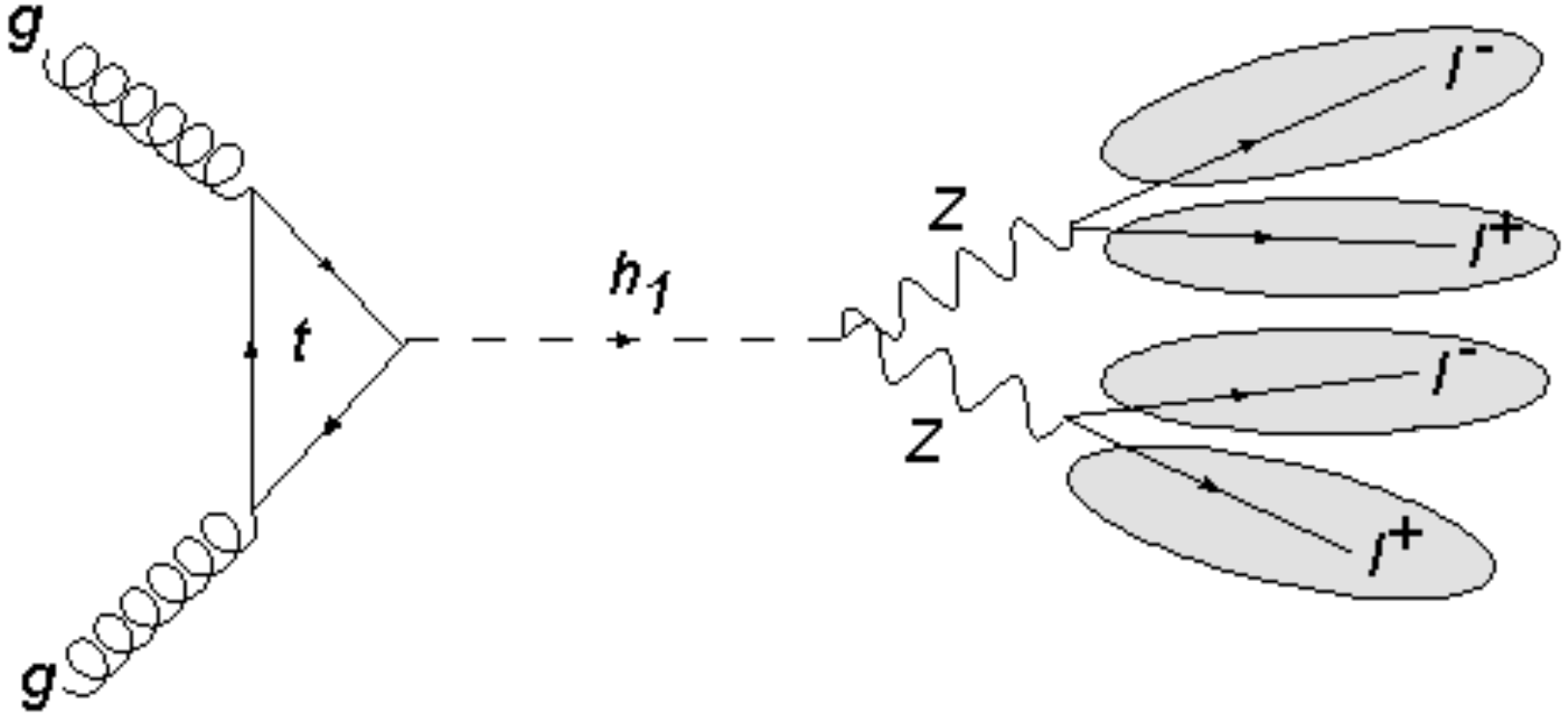}
\includegraphics[width=2.2in]{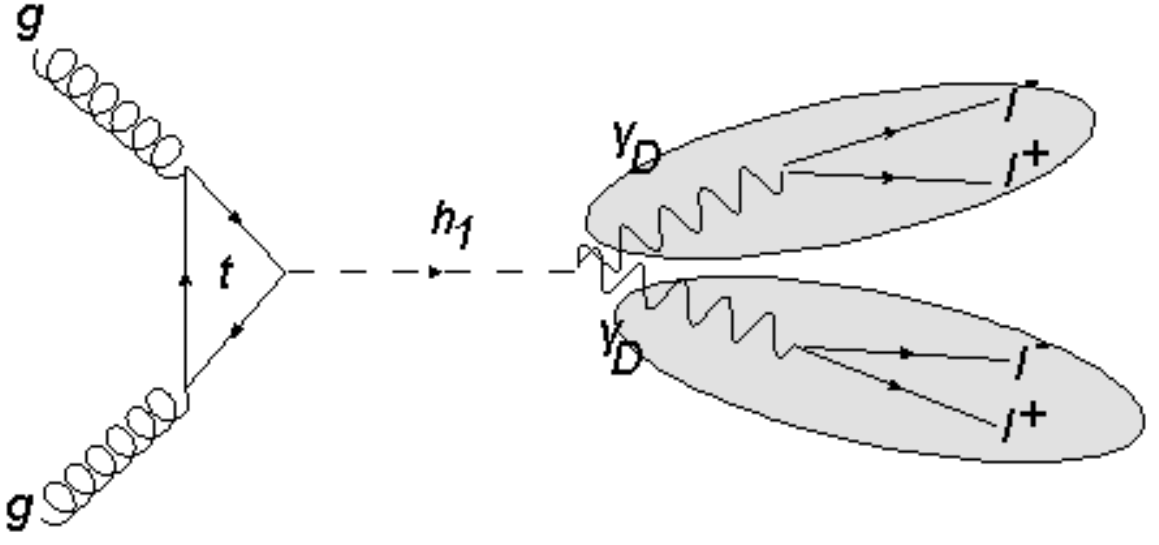}
\caption{\small \label{jets1}
Some topologies of 4 (left) and 2 (right) lepton-jets for process III.
The 4 lepton-jets can also be coming from the SM of process I with $h_1$ replaced by the SM $h$. 
The immediate state of $h_2 h_2$ for the 2 lepton-jets is not shown since the branching ratio 
for $h_2 \to l^+l^-$ is very tiny.
}
\end{figure}

\begin{figure}[h!]
\centering
\includegraphics[width=2.4in]{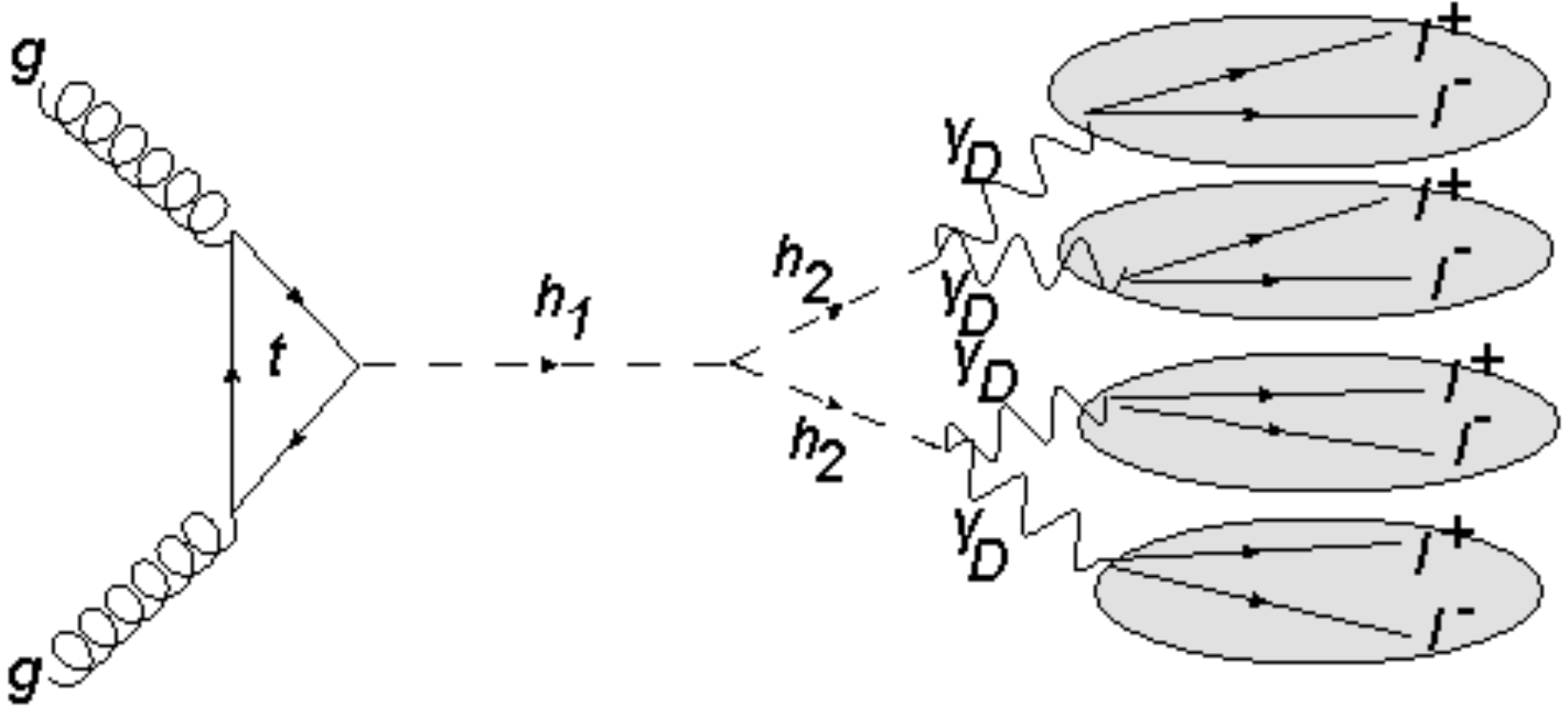}
\includegraphics[width=2.5in]{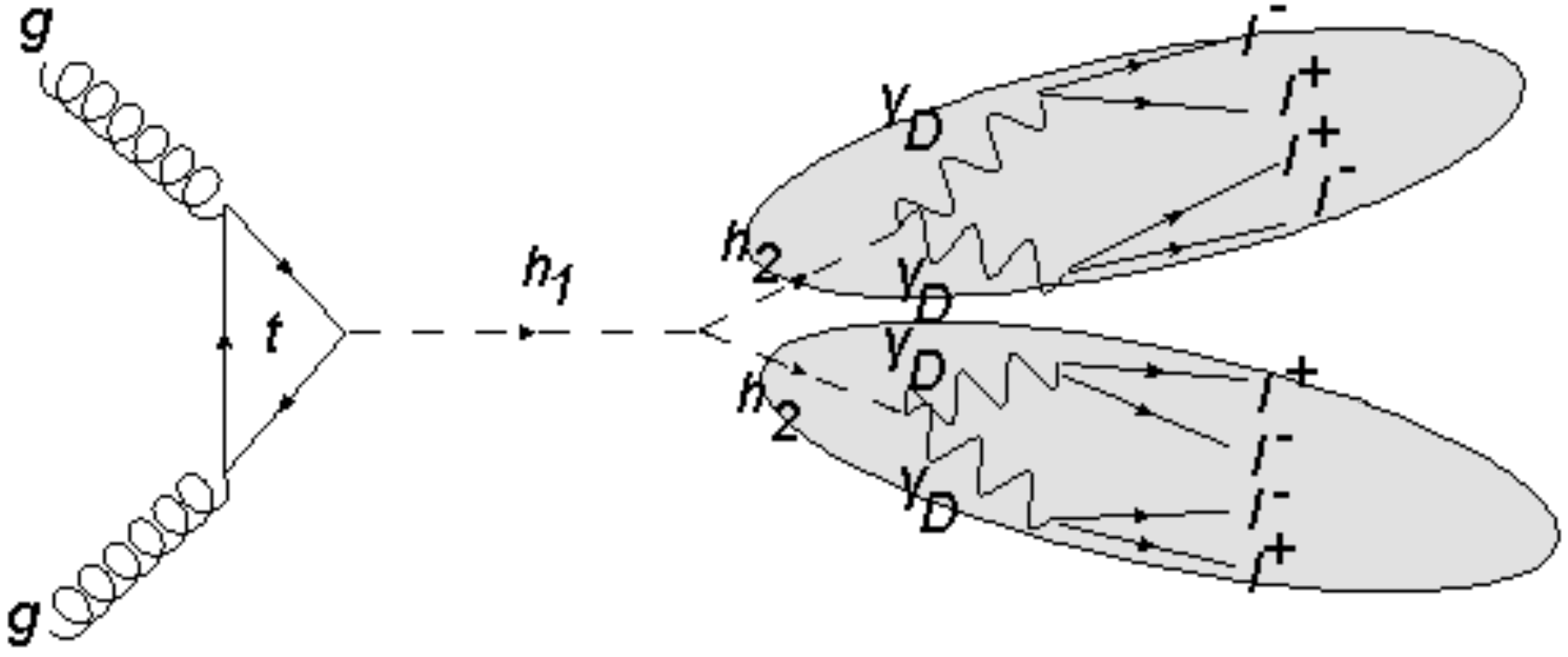}
\caption{\small \label{jets2}
Some topologies of 4 (left) and 2 (right) lepton-jets for process IV.
}
\end{figure}

We will study some collider signatures for the model in this section. In particular, 
we will focus on the 4 lepton-jets and 2 lepton-jets modes in our analysis. 
We consider the following four processes which may lead to signals of multilepton jets
at the LHC: 

\begin{itemize}

\item[(I)] $p p \to h \to Z Z \to l^+ l^- l^+ l^-$

\item[(II)] $p p \to V V \to l^+ l^- l^+ l^-  \; \; \quad (VV \; = \; ZZ, \, \gamma\gamma, \, Z\gamma )$

\item[(III)] $p p \to h_1 \to X X \to l^+ l^- l^+ l^- \; \; \quad (XX \; =  \; ZZ, \, \gamma_D\gamma_D, \, h_2 h_2)$

\item[(IV)] $p p \to h_1 \to h_2 h_2 \to \gamma_D\gamma_D\gamma_D\gamma_D \to l^+ l^- l^+ l^- l^+ l^- l^+ l^-$

\end{itemize}
where $l = e$ or $\mu$.
Processes (I) and (II) are coming entirely from the SM, process (III) can be arise from either SM 
(with modified Higgs-$ZZ$ coupling) or the dark portal (see Fig.~\ref{jets1}), 
and process (IV) is purely from the dark portal (see Fig.~\ref{jets2}).

We compute the matrix elements of these processes using {\sf FeynRules} 
\footnote{We include both gluon and photon fusion $g g \to h_1$ and $\gamma \gamma \to h_1$ 
computed at next-to-leading-order.} \cite{feynrules1,feynrules2}  and {\sf MadGraph} \cite{madgraph}.
We pass these matrix elements to the event generator {\sf MadEvent} \cite{madanalysis} to obtain our 
event samples.  
The set of parton distribution functions used is {\sf CTEQ6L1} \cite{cteq}.

For illustration, we will choose several benchmark points in the dark portal as shown in Table~\ref{table:benchmarks}.
If the kinetic mixing parameter $\epsilon$ is smaller than $10^{-5}$, the dark photon will have a very long lifetime
and it may decay outside the detector.  We will choose it to be $10^{-4}$ as used by previous analyses by theorists  
\cite{Zupan-1} as well as experimentalists \cite{Aad:2013yqp}. 
The mass of dark photon is chosen to be less than 2 GeV in these benchmark points. 
With such relatively low mass the opening angle of the lepton pair from the decay of the dark photon
will be small which may lead to multilepton jets.
Such low mass dark photon may also be desirable for indirect dark matter searches,
since the allowed decay $\gamma_D \to e^+ e^-$ may be used to explain the positron excess 
\cite{Abdo:2009zk,Adriani:2008zr,Aguilar:2013qda},
while $\gamma_D \to p \bar p$ is kinematically disallowed in accord with observation
that the cosmic anti-proton flux is consistent with the background
\cite{Adriani:2008zr,Adriani:2010rc}.
We also choose $\sin^2\alpha = {10^{-3}}$ in consistent with the analysis of the invisible branching ratio of 
$h_1$ in previous section (see also \cite{Clarke:2013aya} and \cite{L3-OPAL,Searches:2001ab}).
At these benchmark points, we see from the last three columns of Table~\ref{table:benchmarks} 
that (1) the invisible decay branching ratio of the SM Higgs 
is consistent with global fit results, (2) the decay of the dark Higgs is almost 100\% into pair of dark photons,
and (3) the branching ratio of the dark photon into light lepton pairs can be as large as 50\%.
Due to the smallness of the two mixing parameters $\alpha$ and $\epsilon$,
the production cross section of $h_1$ at the LHC remains to be very close to its SM value.

%
\begin{table}[t!] 
\caption{Several benchmark points of the dark portal used to calculate the signals of multilepton jets.
($\epsilon = 10^{-4}$ and $\sin ^2 \alpha = 10^{-3}$)} 
\centering 
\begin{tabular}{c  c c c c c c} 
\hline\hline 
Benchmark Point & $g_D$  & $\quad$ $M_{\gamma_D}$ & $m_2$ & Br$_{h_1 \to {\rm Dark Stuff}}$ & Br$_{h_2 \to \gamma_D\gamma_D}$ & Br$_{\gamma_D \to l^+l^-}$\\ [1ex] 
\hline 
A & $\quad$ $0.005$ $\quad$  & $\quad$ $1.5$ & $\quad$ 4 $\quad$ & $\sim16\%$ & $99\%$ & $50\%$ \\ 
B & $\quad$ $0.009$ $\quad$  & $\quad$ $1.8$ & $\quad$ 10 $\quad$ & $\sim20\%$ & $100\%$ & $50\%$ \\ 
C & $\quad$ $0.005$ $\quad$  & $\quad$ $1.5$ & $\quad$ 40 $\quad$ & $\sim15\%$ & $99\%$ & $50\%$ \\
D & $\quad$ $0.005$ $\quad$  & $\quad$ $1.8$ & $\quad$ 40 $\quad$ & $\sim11\%$ & $99\%$ & $50\%$ \\
\hline 
\end{tabular} 
\label{table:benchmarks} 
\end{table} 

For the kinematic cuts for the 2 and 4 lepton-jets, we follow Refs.~\cite{Gopalakrishna:2008dv,Zupan-1} and 
\cite{Aad:2013yqp}. For the basic cuts that we will impose in all processes, we have\\
\noindent
\underline{Basic cuts}: \\
\indent (4 leptons case) $p_{T_l} \geq 20,10,10,10$ GeV, $\quad$ $\quad$ $|\eta_l| < 2.3$;\\
\indent (8 leptons case) $p_{T_l} \geq 20,10,10,10,0,0,0,0$ GeV, $\quad$ $\quad$ $|\eta_l| < 2.3$,\\
where $p_{T_l}$ and $\eta_l$ are the transverse momenta and pseudo-rapidity of the lepton respectively.
On top of the basic cuts, we employ the following lepton-jets cuts\\
\noindent
\underline{4 lepton-jets cuts}:$\quad$ $\Delta R^d_{j_ij_j} > 0.7$,$\quad$ $\Delta R^s_{l_il_j} < 0.2$, 
$\quad$ $M_{\rm Invariant} = M_{h_1} \pm{} 10 $ GeV;

\noindent
\underline{2 lepton-jets cuts}:$\quad$ $\Delta R^d_{j_1j_2} > 0.7$,$\quad$ $\Delta R^s_{l_il_j} < 0.2$, 
$\quad$ $M_{\rm Invariant} = M_{h_1} \pm{} 10 $ GeV.

\begin{figure}[b!]
\centering
\includegraphics[width=2.4in]{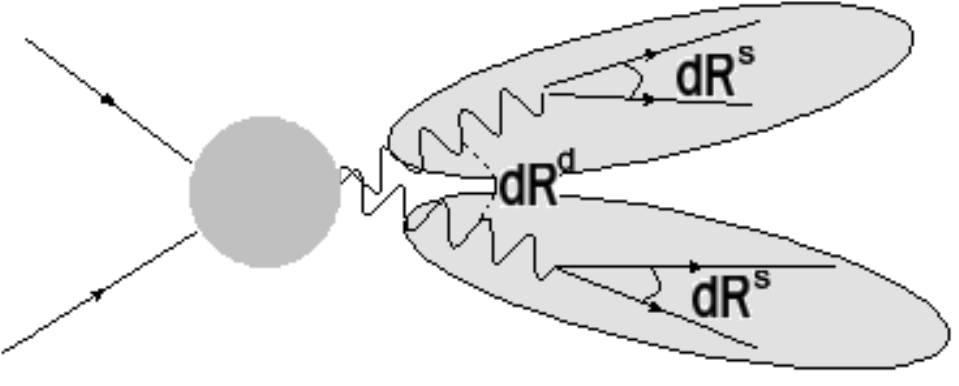}
\includegraphics[width=2.2in]{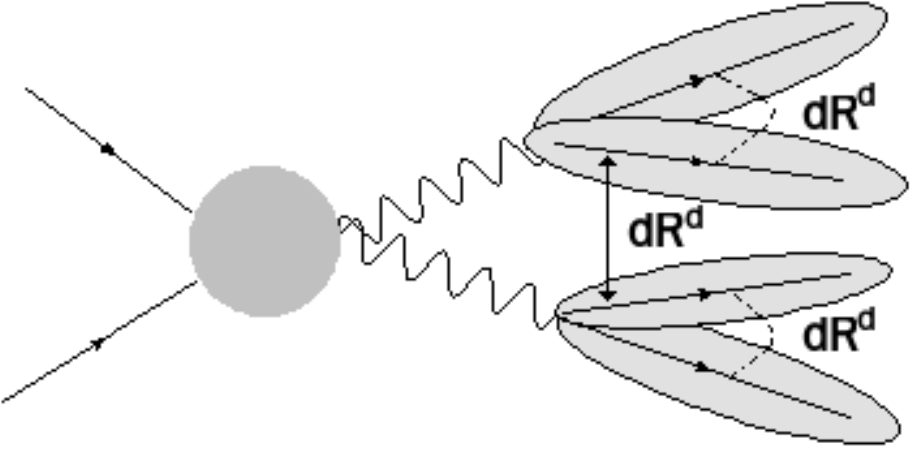}
\caption{\small \label{cuts}
Graphical illustrations for the kinematic cuts on the cone radius $\Delta R$ of final state leptons. 
The 2 and 4 lepton-jets cases are shown in the left and right figures respectively.}
\end{figure}

\noindent
Here $\Delta R^d_{jj}$ denotes the cone radius between two different lepton-jets and 
$\Delta R^s_{ll}$ denotes the cone radius between two different leptons in the same lepton jet, 
as depicted in Fig.~\ref{cuts}. 
$M_{\rm Invariant}$ denotes the invariant mass of all final state particles due to the decay chain of
the SM Higgs boson resonance, give or take 10 GeV from the central value of 126 GeV.

\begin{figure}[b!]
\centering
\includegraphics[width=3.8in]{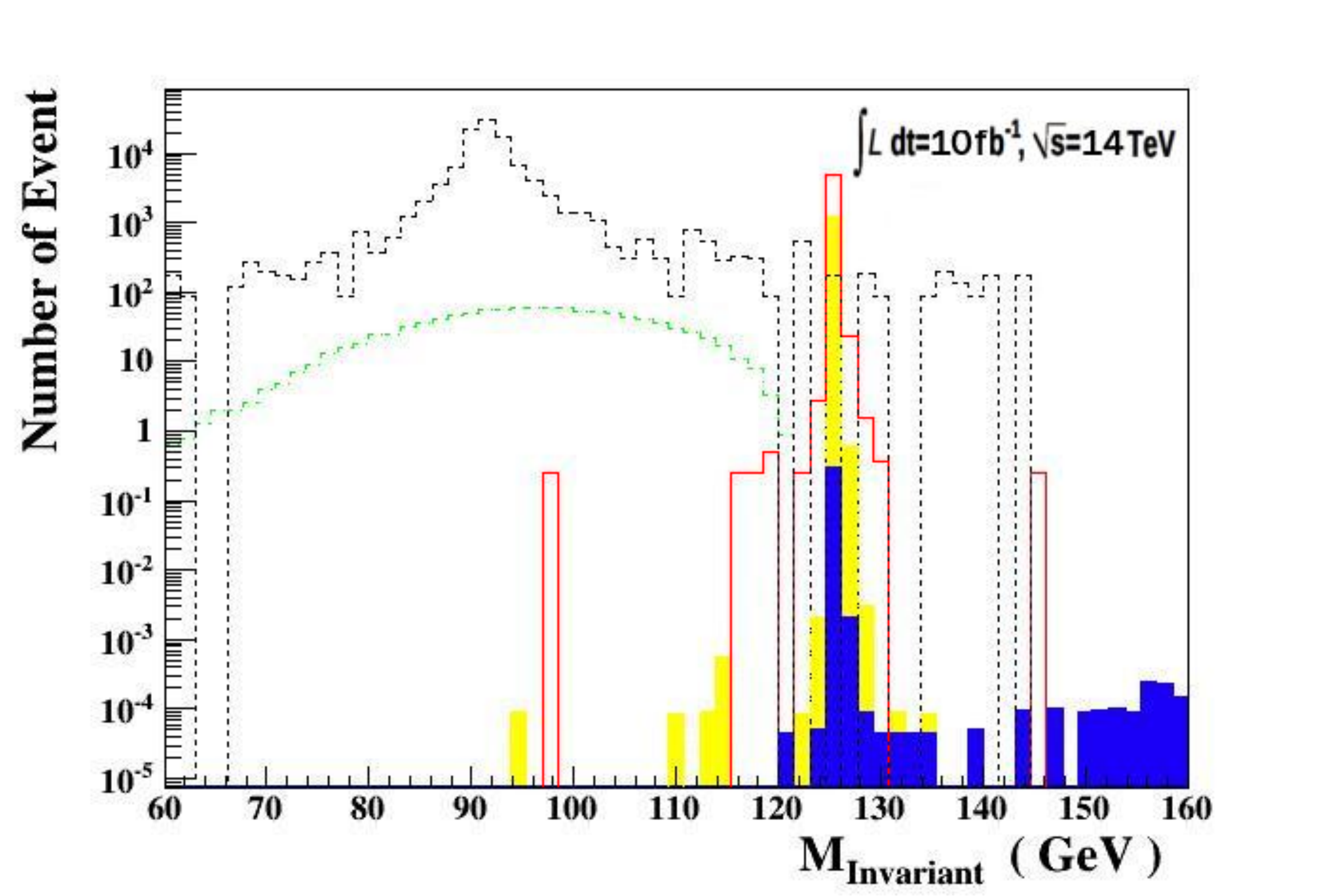}
\caption{\small \label{event}
Number of events versus $M_{\rm Invariant}$ [GeV] with the basic cuts for 
benchmark point B at LHC-14 with a fixed luminosity of 10 fb$^{-1}$.
Histogram of blue strip is for process I, black dash is for process II, 
red solid is for process III, yellow strip is for process IV of 8 leptons, and green dash is for process IV 
of 4 leptons. }
\end{figure}

The number of events versus the total invariant mass $M_{\rm Invariant}$ 
for the four processes I, II, III and IV at the LHC-14 
without any cuts are shown in Fig.~\ref{event} for the benchmark point B.
We can see that before imposing any cuts the number of events around the Higgs boson resonance 
for the two processes III (red) and IV (yellow, 8 leptons) from the dark portal 
can stand above the SM processes of I (blue) and II (black). However away from the 
resonance region, the 4 leptons SM background from process I
(black) is 2 to 3 order of magnitudes above the signals from process IV (green).

We now discuss the impact of imposing the multilepton jets cuts 
on the cross sections. The topologies of imposing the 4 and 2 lepton-jets cuts for processes
III and IV are shown in Figs.~\ref{jets1} and \ref{jets2} respectively.
In Table~\ref{table:xsections}, we show the cross sections of the 4 processes at the LHC-14 
with the basic, 4 and 2 lepton-jets cuts for the 4 benchmark points listed in Table~\ref{table:benchmarks}.
The following statements can be drawn from the results shown in Table~\ref{table:xsections}:
\begin{itemize}

\item
The 4 and 2 lepton jets cuts have strong and different 
impact for the SM processes I and II. 
For process I, since the intermediate state is the $Z$ boson with a relatively high mass,
its decay products can be produced at a relatively large angle with respect to the original $Z$ boson direction.
Thus it favors 4 lepton-jets in the final state (see left diagram in Fig.~\ref{jets1})
and 2 lepton-jets is vanishing small for process I.
On the other hand, SM process II has a cross section of about 700 times larger 
than process I with just the basic cuts imposed. Imposing the 4 and 2 lepton-jets
cuts reduce the cross section of process II by a factor of $4.7 \times 10^{-3}$ and $1.1 \times 10^{-3}$ respectively.
We note that the $ZZ$ intermediate state in process II arises from the tree level parton processes
of quark-quark annihilation while in process I it is connected with the loop-induced gluon fusion mechanism
of Higgs production.

\item
For process III since the dark photon mass is small 
(1.5 GeV for benchmark points A and C, and 1.8 GeV for benchmark points B and D)
the contribution from intermediate state of $\gamma_D \gamma_D$ will give rise mainly to 2 lepton jets 
(see right diagram in Fig.~\ref{jets1}). 
Thus imposing the 4 lepton jets cuts for process III will suppress this intermediate state and only
the contribution from $ZZ$ intermediate state will survive (see left diagram in Fig.~\ref{jets1}).
Since this $ZZ$ contribution is very similar to the SM process I, they should have very similar cross sections
after imposing 4 lepton-jets cuts as clearly seen in Table~\ref{table:xsections}.
On the other hand, imposing 2 lepton-jets cuts will suppress the $ZZ$ intermediate state but 
keep the $\gamma_D \gamma_D$. However, the contribution of $ZZ$ intermediate state for 
process III is negligible. The 2 lepton-jets cross sections of process III are several orders of magnitudes 
larger than the corresponding cross sections of SM process II.

\item
For process IV, with just basic cuts its cross section is about a factor 4 (benchmark points A and B) 
to 5 (benchmark points C and D) smaller than that of process III. However, due to the small mass of the 
dark photon (compared with $Z$ boson mass), one can has either 4 or 2 lepton-jets in the final state.
Imposing the 4 and 2 lepton-jets cuts in addition to the basic cuts for process IV have more nontrivial 
effects on the cross section depending on the benchmark points. For 4 lepton-jets
the cross sections can reach about 3 and 1 femtobarn  for benchmark points C and D respectively.
For 2 lepton-jets, the cross section can reach 2 femtobarn for benchmark point A only.
At these benchmark points, these cross sections are an order of magnitude 
larger than the corresponding cross sections of the SM process II.
Other benchmark points have negligible cross sections for 4 and 2 lepton-jets 
as can be clearly seen in the last column of Table~\ref{table:xsections}. 

\end{itemize}

\begin{table}[ht!]
\caption{Cross sections (in unit of fb) at the LHC-14 for the background processes (I and II) and dark sector processes 
(III and IV) with the basic, 4 and 2 lepton-jets cuts at the 4 benchmark points.} 
\centering
\begin{tabular}{l c r r r r}
\hline\hline
Cuts & Benchmark Point &\multicolumn{4}{c}{$\quad$ I $\quad$ $\quad$ $\quad$ II $\quad$ $\quad$ $\quad$ III $\quad$ $\quad$ $\quad$ $\quad$ IV}
\\ [0.5ex]
\hline
 &A & 0.118 $\quad$ & $70.7$ & 95.3 $\quad$ & 23.2  \\[-0.4ex]
\raisebox{2ex}{Basic}
& B &0.118 $\quad$ & $70.7$ & 204 $\quad$ & 45.8  \\[0.2ex]
 &C &0.118 $\quad$ & $70.7$ & 96.7 $\quad$ & 19.2  \\[0.4ex]
  &D &0.118 $\quad$ & $70.7$ & 68.3 $\quad$ & 13.1  \\[0.4ex]
\hline
 &A & 9.63$\times10^{-3}$ & $\quad$ $0.337$ & $\quad$ 9.86$\times10^{-3}$ & $\quad$ $\leq10^{-10}$  \\[-0.8ex]
\raisebox{2ex}{Basic + 4 Lepton-Jets}
& B &9.63$\times10^{-3}$ & $\quad$ $0.337$ & $\quad$ 9.80$\times10^{-3}$ & $\quad$ $\leq10^{-10}$  \\[0.4ex]
&C &9.63$\times10^{-3}$ & $\quad$ $0.337$ & $\quad$ 9.93$\times10^{-3}$ & $\quad$ 3.05  \\[0.8ex]
&D &9.63$\times10^{-3}$ & $\quad$ $0.337$ & $\quad$ 9.84$\times10^{-3}$ & $\quad$ 0.92  \\[0.8ex]
\hline
 &A & $\leq10^{-10}$ & $0.08$ & $\quad$ 95.3 $\quad$ & $1.75$  \\[-0.8ex]
\raisebox{2ex}{Basic + 2 Lepton-Jets}
& B & $\leq10^{-10}$ & $0.08$ & $\quad$ 201 $\quad$ & $\leq10^{-10}$   \\[0.4ex]
 &C & $\leq10^{-10}$ & $0.08$ & $\quad$ 95.8 $\quad$ & $\leq10^{-10}$  \\[0.8ex]
 &D & $\leq10^{-10}$ & $0.08$ & $\quad$ 68.2 $\quad$ & $\leq10^{-10}$  \\[0.8ex]
\hline
\end{tabular}
\label{table:xsections}
\end{table}
%


\section{Conclusions}

We have studied a simple extension of the SM by adding a dark sector described by 
the original Abelian Higgs model. The communication between the visible sector and the dark
sector is due to the mixing between the SM and dark Higgses and/or mixing between the
SM and dark photons. We study various non-standard decay modes of the heavier Higgs $h_1$ in this model,
identified as the 126 GeV new boson observed recently at the LHC. 
Multilepton modes in the final states of this heavier Higgs decay are possible. For the case of
$h_1 \to h_2 h_2$ followed by $h_2 \to \gamma_D \gamma_D$ and $\gamma_D \to \bar l l$,
there could be eight leptons in the final states. The three body process $h_1 \to h_2 \gamma_D \gamma_D$
is found to be significant and could lead to eight leptons final state as well. On the other hand, 
the other three body process $h_1 \to h_2 h_2 h_2$ has an insignificant branching ratio; otherwise,
it would lead up to a even more spectacular twelve leptons final state.
The signals of 4 and 2 lepton-jets in this model are already quite unique and spectacular.
We show that there are parameter space in this simple dark portal model satisfying 
the current constraint of the non-standard decay width of the 126 GeV Higgs and
may give rise to interesting signals of multilepton jets at the LHC-14.
Experiments at the LHC should therefore search for 
multilepton modes in the Higgs decay in order to probe for the possible existence
of a $U(1)_D$ dark sector governed by the original Abelian Higgs model.



\section*{Acknowledgment}

EM would like to thank the hospitality of the Institute of Physics, Academia Sinica, Taipei, Taiwan. 
We would like to thank Chuan-Ren Chen, Sunghoon Jung, Zuowei Liu and Olivier Mattelaer for many 
useful communications.
This work is supported in part by the U.~S.~Department of Energy under 
Grant No.~DE-SC0008541 and by National Science Council of Taiwan under grant 
101-2112-M-001-005-MY3.


\bibliographystyle{unsrt}

\end{document}